\documentclass[useAMS, usenatbib]{mn2e}
\usepackage{subfigure}
\usepackage{ulem}
\usepackage{color}
\usepackage{graphicx}
\usepackage{times} 
\usepackage{amssymb}
\usepackage{amsmath}
\usepackage{lscape}
\usepackage{url}
\usepackage{multirow}
\usepackage{longtable}
\usepackage[english]{babel}
\usepackage{float}
\newif\ifAMStwofonts
\AMStwofontstrue

\def\btheta{\vec{\theta}}
\newcommand{\velocitydispersion}{$\sigma_0$}
\newcommand{\coreradius}{$r_{core}$}
\newcommand{\cutradius}{$r_{cut}$}
\newcommand{\velocitydispersionLstar}{$\sigma^{*}_0$}
\newcommand{\coreradiusLstar}{$r^{*}_{core}$}
\newcommand{\cutradiusLstar}{$r^{*}_{cut}$}
\newcommand{\bcga}{BCG-A}
\newcommand{\bcgb}{BCG-B}
\newcommand{\solMass}{\ensuremath{\rmn{M}_{\odot}}}
\newcommand{\lenstool}{{\sevensize LENSTOOL}}
\newcommand{\glafic}{{\sevensize GLAFIC}}
\newcommand{\sourceextractor}{{\sevensize SEXTRACTOR}}

\title[Strong Lensing by Abell 2146]{The Mass Distribution of the Unusual Merging Cluster Abell 2146 from Strong Lensing}
\author[J. E. Coleman, et. al.]{
	Joseph E. Coleman$^{1}$\thanks{E-mail:joseph.coleman@utdallas.edu}, 
	Lindsay J. King$^{1}$, 
	Masamune Oguri$^{2,3,4}$, \newauthor{}
	Helen R. Russell$^{5}$, 
    Rebecca E. A. Canning$^{6,7}$,
	Adrienne Leonard$^{8}$,  
	Rebecca Santana$^{9}$, \newauthor{}
	Jacob A. White$^{1,10}$,
	Stefi A. Baum$^{11}$, 
    Douglas I. Clowe$^{9}$,
    Alastair Edge$^{12}$, \newauthor{}
	Andrew C. Fabian$^{5}$,
	Brian R. McNamara$^{13}$,
    and
	Christopher P. O'Dea$^{11}$ 
\\
$^{1}$Department of Physics, The University of Texas at Dallas, 800 W. Campbell Road, Richardson, TX 75080, USA\\
$^{2}$Department of Physics, University of Tokyo, 7-3-1 Hongo, Bunkyo-ku, Tokyo 113-0033, Japan \\
$^{3}$Kavli Institute for the Physics and Mathematics of the Universe (Kavli IPMU, WPI), University of Tokyo, Chiba 277-8583, Japan \\
$^{4}$Research Center for the Early Universe, The University of Tokyo, 7-3-1 Hongo, Bunkyo-ku, Tokyo 113-0033, Japan \\
$^{5}$Institute of Astronomy, University of Cambridge, Madingley Road, Cambridge CB3 0HA, UK \\
$^{6}$Kavli Institute for Particle Astrophysics and Cosmology, Stanford University, 452 Lomita Mall, Stanford, CA 94305-4085, USA \\
$^{7}$Department of Physics, Stanford University, 382 Via Pueblo Mall, Stanford, CA 94305-4060, USA \\
$^{8}$Department of Physics and Astronomy, University College London, Gower Place, London WC1E 6BT, UK \\
$^{9}$Department of Physics and Astronomy, Ohio University, Clippinger Labs 251B, Athens, OH 45701, USA \\
$^{10}$Department of Physics and Astronomy, University of British Columbia,  6224 Agricultural Rd. Vancouver BC V6T 1Z1, Canada \\
$^{11}$University of Manitoba, Department of Physics and Astronomy, Winnipeg, MB R3T 2N2, Canada \\
$^{12}$Department of Physics, Durham University, Durham DH1 3LE \\
$^{13}$Department of Physics and Astronomy, University of Waterloo, Waterloo, N2L 3G1, Canada}

\begin{document}

\date{Accepted 2016 September 21. Received 2016 September 16; in original form 2016 May 10}

\pagerange{\pageref{firstpage}--\pageref{lastpage}} \pubyear{2016}

\maketitle

\label{firstpage}

\begin{abstract}
Abell 2146 consists of two galaxy clusters that have recently collided close to the plane of the sky, and it is unique in showing two large shocks on {\it Chandra X-ray Observatory} images. With an early stage merger, shortly after first core passage, one would expect the cluster galaxies and the dark matter to be leading the X-ray emitting plasma. In this regard, the cluster Abell 2146-A is very unusual in that the X-ray cool core appears to lead, rather than lag, the Brightest Cluster Galaxy (BCG) in their trajectories. Here we present a strong lensing analysis of multiple image systems identified on {\it Hubble Space Telescope} images.  In particular, we focus on the distribution of mass in Abell 2146-A in order to determine the centroid of the dark matter halo.  We use object colours and morphologies to identify multiple image systems; very conservatively, four of these systems are used as constraints on a lens mass model.  We find that the centroid of the dark matter halo, constrained using the strongly lensed features, is coincident with the BCG, with an offset of  $\approx$ 2 kpc between the centres of the dark matter halo and the BCG. Thus from the strong lensing model, the X-ray cool core also leads the centroid of the dark matter in Abell 2146-A, with an offset of $\approx$ 30 kpc. 
\end{abstract}

\begin{keywords}
gravitational lensing: strong; galaxies: clusters: general; galaxies: clusters: individual: Abell 2146
\end{keywords}

\section{Introduction}

\begin{figure*}
	\includegraphics[width=\textwidth]{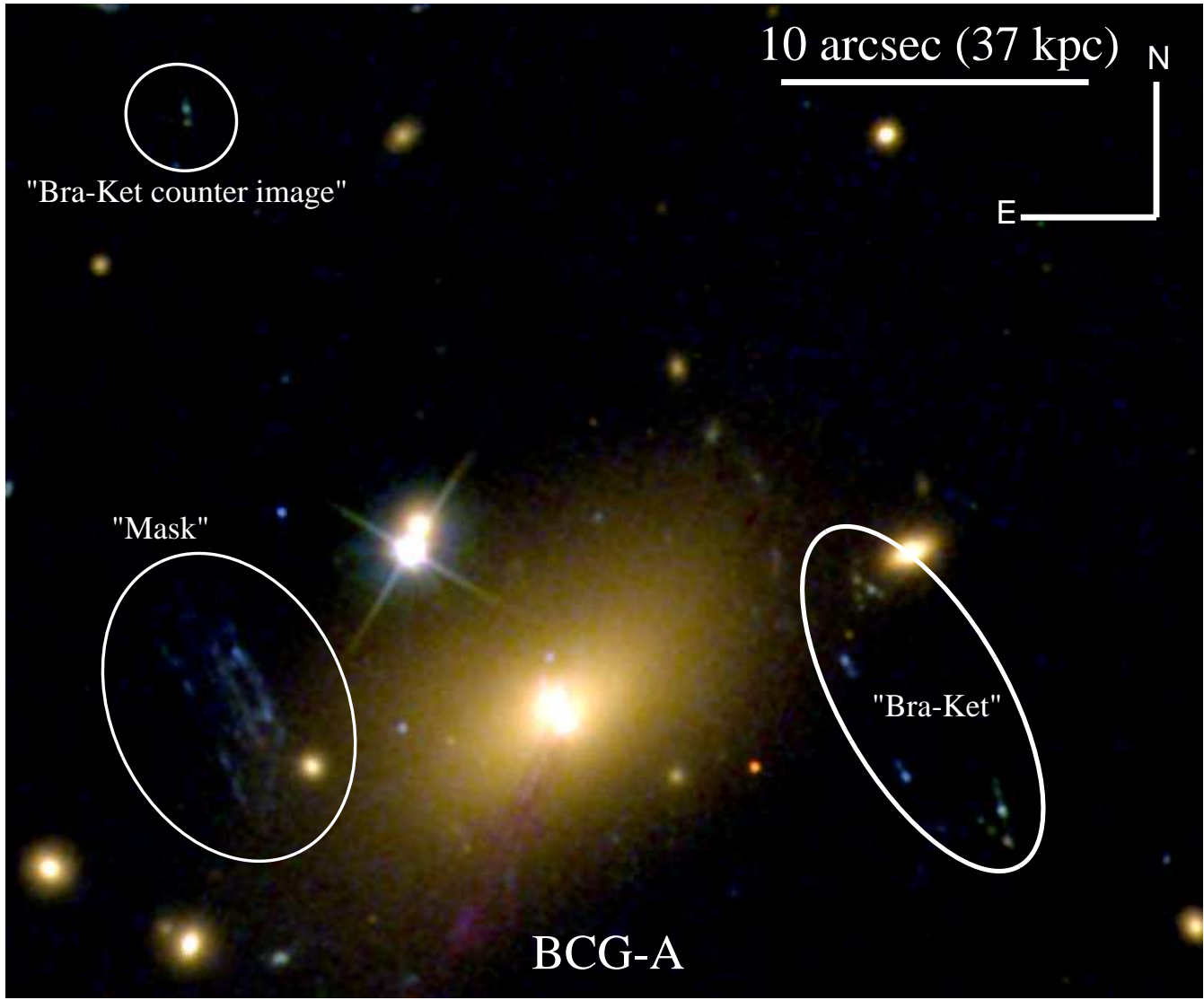}
	\caption{A false colour view of a {\it Hubble Space Telescope} of the brightest cluster galaxy in cluster A. There are two prominent strong lensing features with distinct bilateral symmetry, the `mask' system and the `bra-ket' system.  The mask system has two emission knots that look like eyes.  The bra-ket system resemble the symbols $<$ and $>$. South East of the centre of BCG-A is jet of gas from the active galactic nuclei.}
	\label{figMaskAndBraKet}
\end{figure*}

The cluster system Abell 2146 was first discovered to consist of two massive clusters undergoing a major merger by \citet{russel2010abel2146}, with an appearance on {\it Chandra X-ray Observatory} images reminiscent of the Bullet Cluster \cite[e.g.][]{bullet3}. On X-ray images, the system is unique in presenting two large shocks of Mach number $\sim 2$ \citep{russel2010abel2146,russellxray1}, indicative of a relatively recent merger between two clusters more similar in mass than those in the Bullet Cluster. Estimates from X-ray analysis \citep{russel2010abel2146,russellxray1} and dynamical analysis \citep{white2146dynamics} are consistent with a merger observed about 0.1-0.2 Gyr after first core passage, recent on the dynamical time scales of clusters. Abell 2146 holds great promise for investigating the transport processes in the plasma in cluster environments \citep{russellxray1}.

Dark matter accounts for about 85\% of the mass of galaxy clusters. Most of the baryonic mass, accounting for about 15\% of the total gravitating mass,  is hot X-ray emitting plasma, and at most a few percent of the total mass resides in the stellar components of galaxies. Major mergers of galaxy clusters occurring close to the plane of the sky are very rare events, and their importance in cosmology has been highlighted by the findings from the first such system to be discovered, the Bullet Cluster (e.g. \citealt{bullet3, markevitch2007}). 

When clusters collide, the clouds of hot plasma are slowed down by ram pressure, whereas the galaxies are essentially collisionless and affected mainly by tidal interactions. Dark matter also does not have a large cross-section for interaction (e.g. \citealt{bullet4, randall2008,Harvey2015darkmattercrosssection}), so shortly after collision the major concentrations of galaxies and the dark matter are expected to lead the plasma clouds (e.g. \citealt{bullet3}). A major merger thus results in a dissociation between the plasma clouds, galaxies and dark matter, the specifics of which depends on the cluster properties and merger geometry.

The X-ray cool core of the cluster component Abell 2146-A \citep{russel2010abel2146,russellxray1} is offset from the Brightest Cluster Galaxy (BCG) seen on {\it Hubble Space Telescope} images \citep{king2146weak,canning2012_a2146}, but it leads rather than lags the BCG in their trajectories. At a later stage in a merger, a gravitational slingshot that causes the plasma to overtake the dark matter and galaxies is possible \citep{hallman2004} but the merger would have to be seen a factor of several times later since first core passage for this explanation to be dynamically viable \citep{russellxray1}. Weak lensing mass reconstruction using the distorted shapes of background galaxies on {\it Hubble Space Telescope} images is consistent with the peak in the dark matter in Abell 2146-A being offset from the X-ray cool core, but the resolution of the mass map is too low to draw a statistically robust conclusion. The galaxies in Abell 2146-B are located ahead of the peak in the plasma density, with the BCG being almost coincident with one of the X-ray shocks. 

\begin{figure*}
	\includegraphics[width=\textwidth]{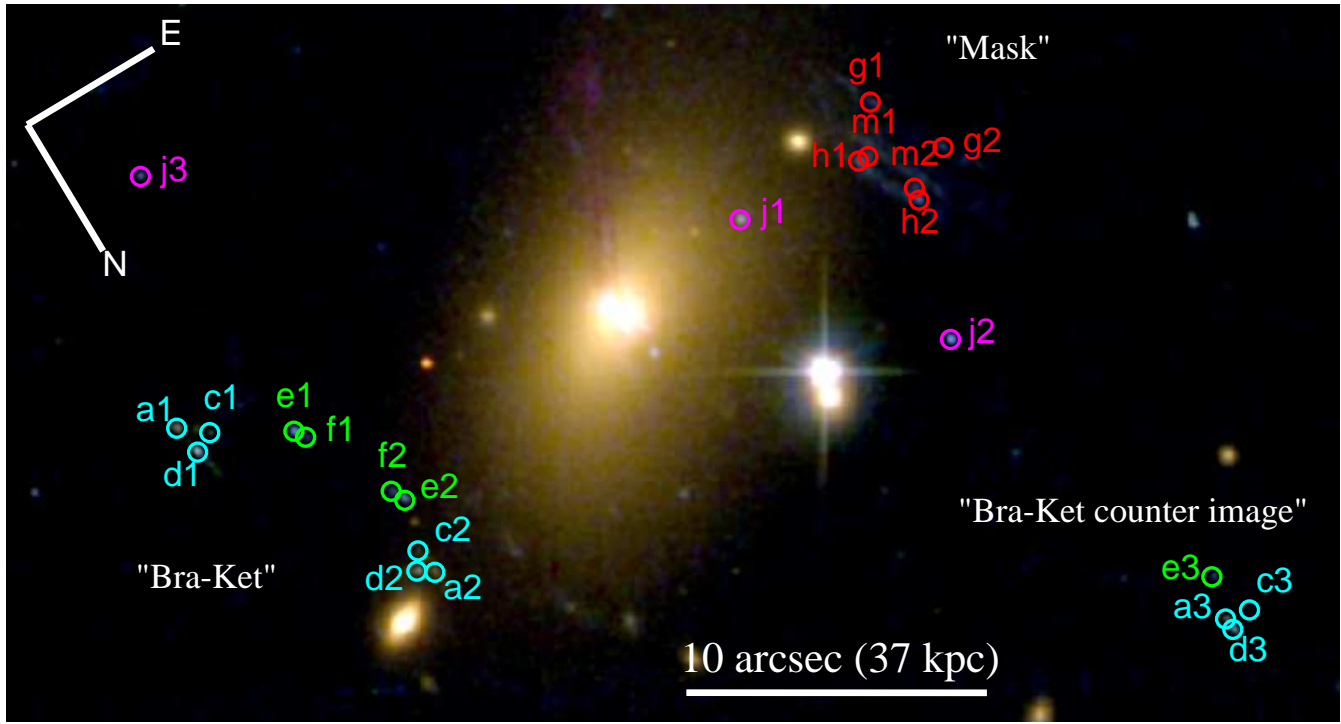}
	\caption{Note the rotation; north is towards the bottom right and east is towards the top right. Multiple image systems used as constraints in the vicinity of \bcga{}. Identical objects have the same letter prefix, e.g. `a1,' `a2,' and `a3.'  The colour grouping indicates the objects considered to be at the same redshift. Objects `a,' `c,' and `d' are in cyan.  Objects `e' and `f' are in green.  Object `j' is in magenta. Objects `g,' `h,' and `m' are in red.  For a close-up view of the systems, see Figures~\ref{figSamuraiMask} and \ref{figBraket}.  The set a, c, and d was statically assigned a redshift of z = 2.0.  Other systems were assumed to have a flat prior for the redshift in the range $z = 0.3$ to $z = 3.0$.  }
	\label{figMultipeImages}
\end{figure*}

Here our aim is to determine the centroid of the dark matter in Abell 2146-A, in order to establish the spatial location with respect to the X-ray cool core, and hence whether it  also lags behind the X-ray cool core in its trajectory. We concentrate on modelling the mass around Abell 2146-A using newly-identified multiple image systems to construct the first strong lensing mass model of the system. Strong lensing analysis offers a higher resolution view of the mass around the BCG than obtained with weak lensing. In order to construct the strong lensing mass model we identify new candidate multiple image systems on {\it HST} images, see Figures~\ref{figMaskAndBraKet} and \ref{figMultipeImages}, on the basis of colours and morphologies. Since we do not have spectroscopic or accurate photometric redshift estimates for the candidate multiple images, we adopt a very conservative approach in our threshold for using a candidate system as a constraint.  

In \S~\ref{SecObsDataReduction}, we discuss the observations and data reduction.  In \S~\ref{SecLensModel}, we discuss the software used and the components of a lens model.  In \S~\ref{SecProcedure} we discuss the procedure used to build the model, including a technique for determining multiple image systems, cluster member selection, model constraints, and computation.  In \S~\ref{SecResults}, we discuss the results of the lens models. 
 
The flat cosmology assumed has $H_{\rmn{0}} = \rmn{70\,km\,s^{-1} Mpc^{-1}}$, $\Omega_{\Lambda} = \rmn{0.7}$, and $\Omega_{m} = \rmn{0.3}$.  Magnitudes used throughout are in the AB system.  At the redshift of z = 0.2323 \citep{white2146dynamics} and given this cosmology, $1$ arcsec = $3.702$ kpc.

For an overview of strong gravitational lensing, see \citet*{Schneider1995book} and \citet{SchneiderSaasFee2006}.

\section{Observations and Data Reduction} \label{SecObsDataReduction} 

The merging cluster Abell 2146 was observed with the {\it Hubble Space Telescope} (HST) on 2013 June 3 and June 6 (HST Cycle 20 proposal 12871, PI:King).  A total of 8 orbits of imaging data was obtained with the ACS/WFC camera.  There were two pointings in each of the f435W and f606W filters, and four pointings in the f814W filter. The pixel scale of the images is $\sim0.05 \farcs $ The reduced fits files were the same as those used for the weak lensing analysis by \citet{king2146weak} which should be referred to for additional details.  A catalogue of objects was created with the freely available software \sourceextractor{} \citep{sourceextractor1996}.  

The position of \bcga{} is the barycentre position as calculated by \sourceextractor{} \citep{sourceextractor1996}.  This location corresponds to the peak in the f814W filter within an error of 1 pixel. The full width half maximum of \bcga{} in the f814W and f606W filters is on the order of 7 pixels or $3.5$ arcseconds.

\begin{figure*}
	\includegraphics[width=0.89\textwidth]{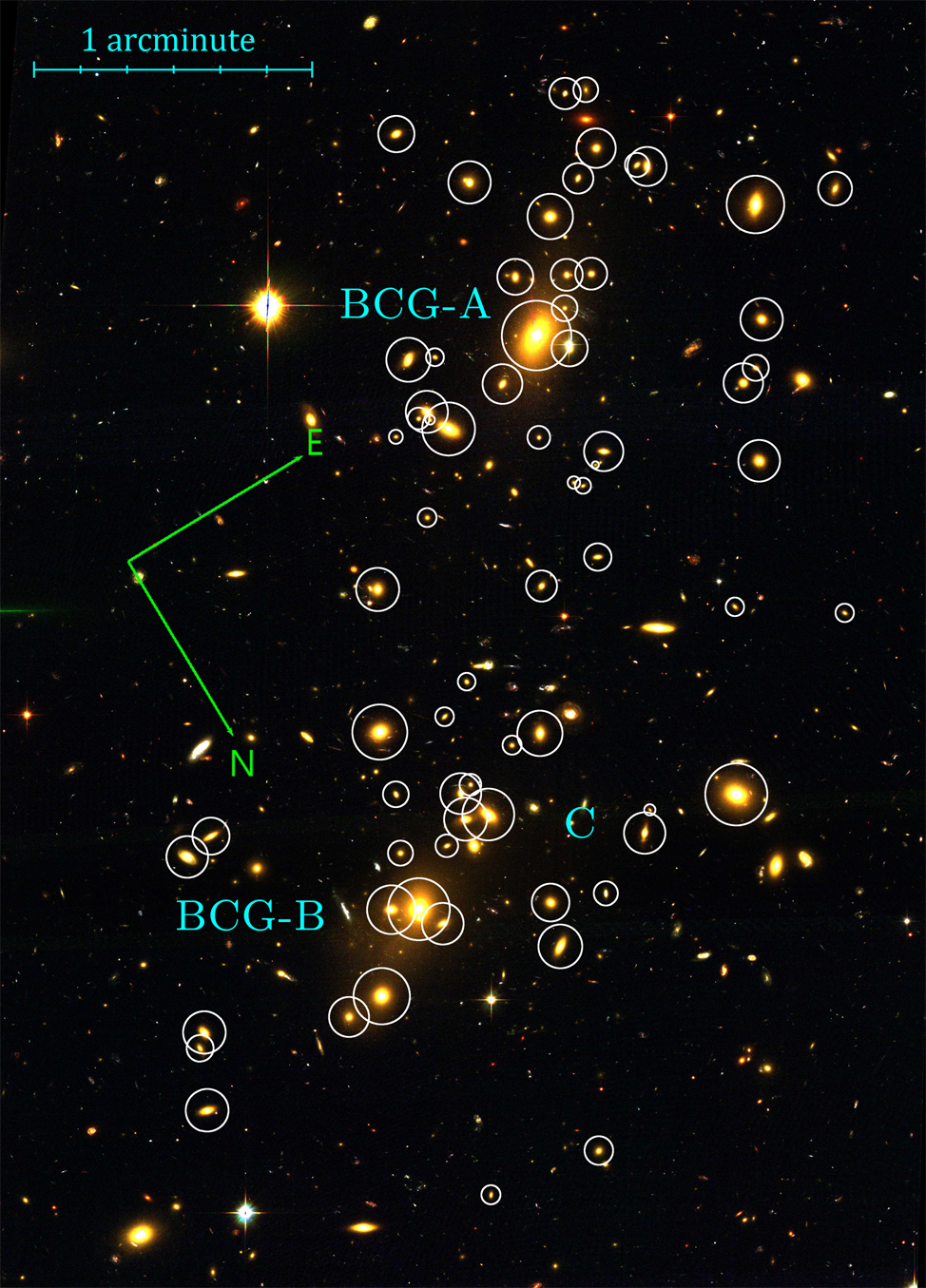}
	\caption{A false colour picture of Abell 2146. The f814W filter was mapped to the colour red, f606W filter was mapped to green, and f435W was mapped to blue. The circles indicate cluster members that are modelled as PIEMD perturbers.  BCG-A and BCG-B indicate the brightest cluster galaxies in each of the respective clusters.  The location of point C is where a mass overdensity appears in the weak lensing analysis \citep{king2146weak}. Circle size has no meaning.}
	\label{figA2146}
\end{figure*}
	
\section{Lens Models} \label{SecLensModel}
The freely available software \lenstool{}\footnote{http://projets.lam.fr/projects/lenstool/wiki} (\citealt{lenstool1996}, \citealt{lenstool2007}), version 6.8, was used for modelling Abell 2146.  \lenstool{} uses Bayesian statistics combined with Markov Chain Monte Carlo techniques to fit a parametrized\footnote{Also known as `Light Traces Matter' or LTM.} lens mass model to gravitational lensing data.  Cluster scale mass density profiles were used to model dark matter components. These density profiles have various parameters that define them, detailed below, and to which some prior is applied. \lenstool{} version 6.8 supports either flat or Gaussian priors on model parameters. Galaxy scale haloes were modelled as perturbers to the cluster scale halo.  The two BCGs were also modelled as perturbers.  Typically, they all have the same type of density profile and have certain parameters scaled collectively, as detailed below. 

The dark matter haloes of Abell 2146 were modelled using a Navarro, Frenk, and White mass density profile \citep*[][hereafter NFW]{NFW1996}. To allow for departure from spherical symmetry, we used the elliptical NFW mass profile implemented in \lenstool{} (Golse \& Kneib 2002).  Any mention of a NFW profile should be interpreted as the elliptical NFW profile. Individual galaxies in the cluster are modelled together as a set of perturbers to the cluster scale NFW halo.  Collectively the perturbers were modelled as pseudo-isothermal elliptical mass distributions, (\citealt{Limousin2007}, \citealt{dPIE2007}\footnote{In \citealt{dPIE2007}, the profile is referred to as dual Pseudo Isothermal Elliptical Mass Distribution.}), hereafter PIEMD.

The parameters used to describe the NFW profile are position ($x, y$), elliptici-y $\epsilon = (a^2 - b^2) / (a^2 + b^2)$ of the projected mass where $a$ and $b$ are the semi-major and semi-minor axes describing the elliptical isodensity contours, $\theta$ which describes the orientation angle of the ellipse, and NFW-specific parameters of mass concentration $c$ and scale radius $R_{s}$.  With the concentration and scale radius related by $c R_{s} = R_{200}$, when a mass $M_{200}$ (the mass contained inside the radius $R_{200}$ where the mean density is 200 times the critical density at the redshift of the halo) is specified, this reduces by one the degrees of freedom of the model.  The NFW profile is truncated at the virial radius.  

The PIEMD perturbers are described by position $(x,y)$, ellipticity $\epsilon$, angle $\theta$, and three PIEMD-specific parameters, velocity dispersion \velocitydispersion{}, cut radius \cutradius{}, and core radius \coreradius{}.  The parameters  \velocitydispersion{}, \cutradius{}, and \coreradius{} are scaled relative to the parameters of an $\rmn{L^*}$ galaxy at the redshift of the cluster; see \citet{lenstool2007} for details. The values used for an $\rmn{L^*}$ galaxy at the redshift of the cluster are given in Table~\ref{tblLstar}. The values for $\rmn{M^*}$, \velocitydispersionLstar{}, \cutradiusLstar{} were adopted from one of the CLASH \citep{Postman2012CLASH} mass models by Zitrin\footnote{http://archive.stsci.edu/missions/hlsp/\\clash/rxj2129/models/zitrin/nfw/v1/}, with his technique explained in \citet{zitrin2013}.  Both \velocitydispersionLstar{} and \cutradiusLstar{} have Gaussian priors.  The core radius \coreradiusLstar{} is a fixed parameter, typical of an $\rmn{L^*}$ galaxy. 

\begin{table}
	\caption{Parameters of an $\rmn{L^*}$ elliptical galaxy used in scaling relations. The parameter $\rmn{m^{*}_0}$ is apparent magnitude, \velocitydispersionLstar{} is velocity dispersion, \cutradiusLstar{} is cut radius, and  \coreradiusLstar{} is core radius.  The core radius is a fixed value.}
	\label{tblLstar}
	\begin{tabular}{l|l}
	\hline  
	$\rmn{m^{*}_0}$			 &  18.070  \\
	\velocitydispersionLstar{} ($\rmn{km s^{-1}}$)  &  99.2 $\pm$ 80.0 \\
	\cutradiusLstar{}			 (kpc) &  49.3 $\pm$ 43.0 \\
	\coreradiusLstar{}		 (kpc) &  [0.15] \\
	\hline
	\end{tabular}
\end{table}

As a consistency check, additional models were created with the software package \glafic{} \citep{glafic_paper2010} as further discussed in the Results and Discussion Section~\ref{SecResults}. Optimization of lens models with \glafic{} used a downhill simplex method \citep{NumericalRecipies}.  As was done with \lenstool{}, the cluster scale dark matter components were modelled with elliptical NFW profiles.  The galaxy components were modelled collectively as perturbers consisting of pseudo-Jaffe ellipsoids \citep{jaffe1983, keeton2001astro0102341}.  In \glafic{}, the ellipticity used is $\epsilon = 1 - b/a$, where $a$ and $b$ are the semimajor and semiminor axes respectively. 

Models in \lenstool{} and \glafic{} had the same constraints and free parameters. 

\section{Procedure} \label{SecProcedure}

\subsection{Cluster Member Selection}

\begin{figure}
	\includegraphics[width=84mm]{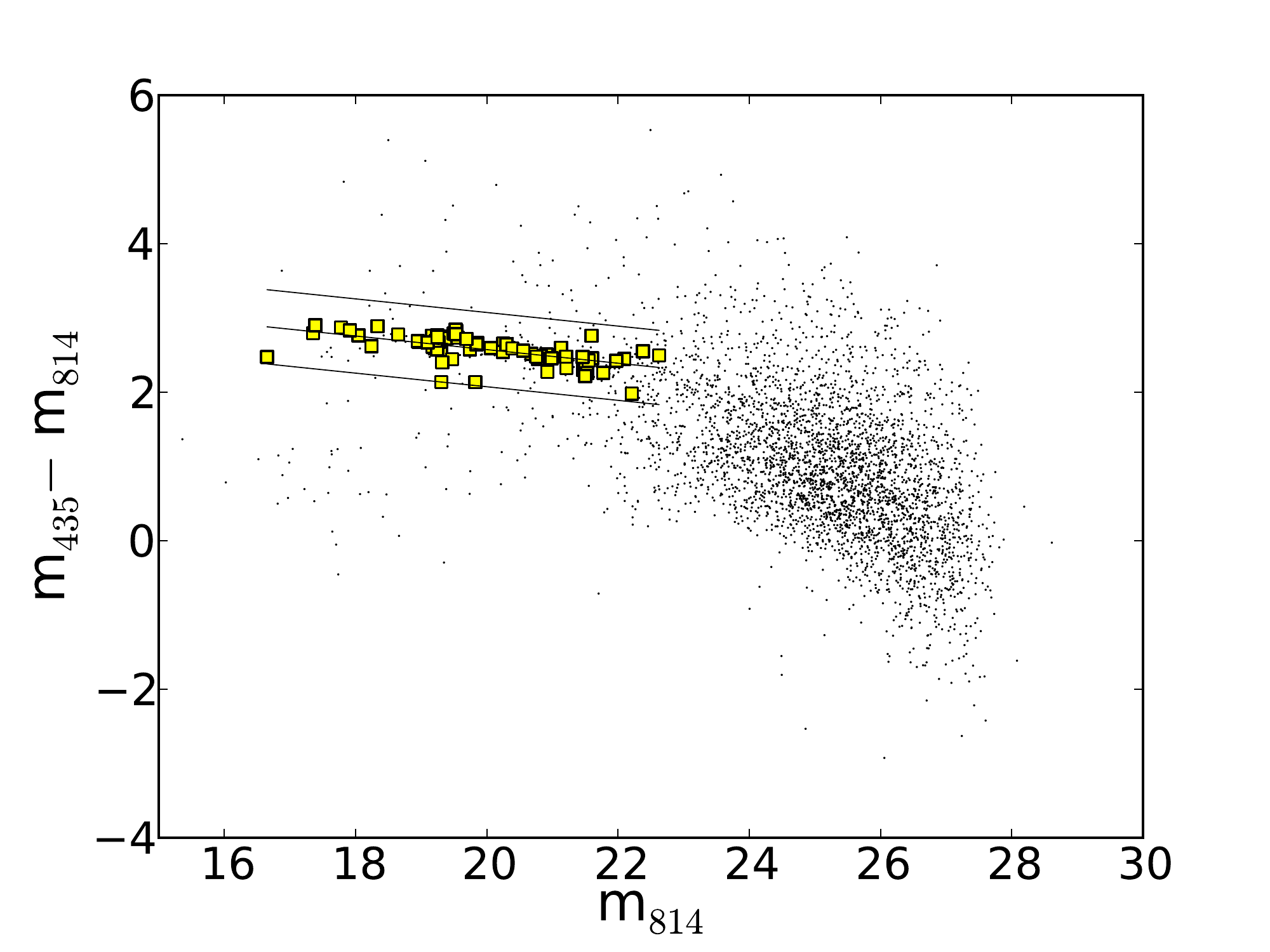}
	\includegraphics[width=84mm]{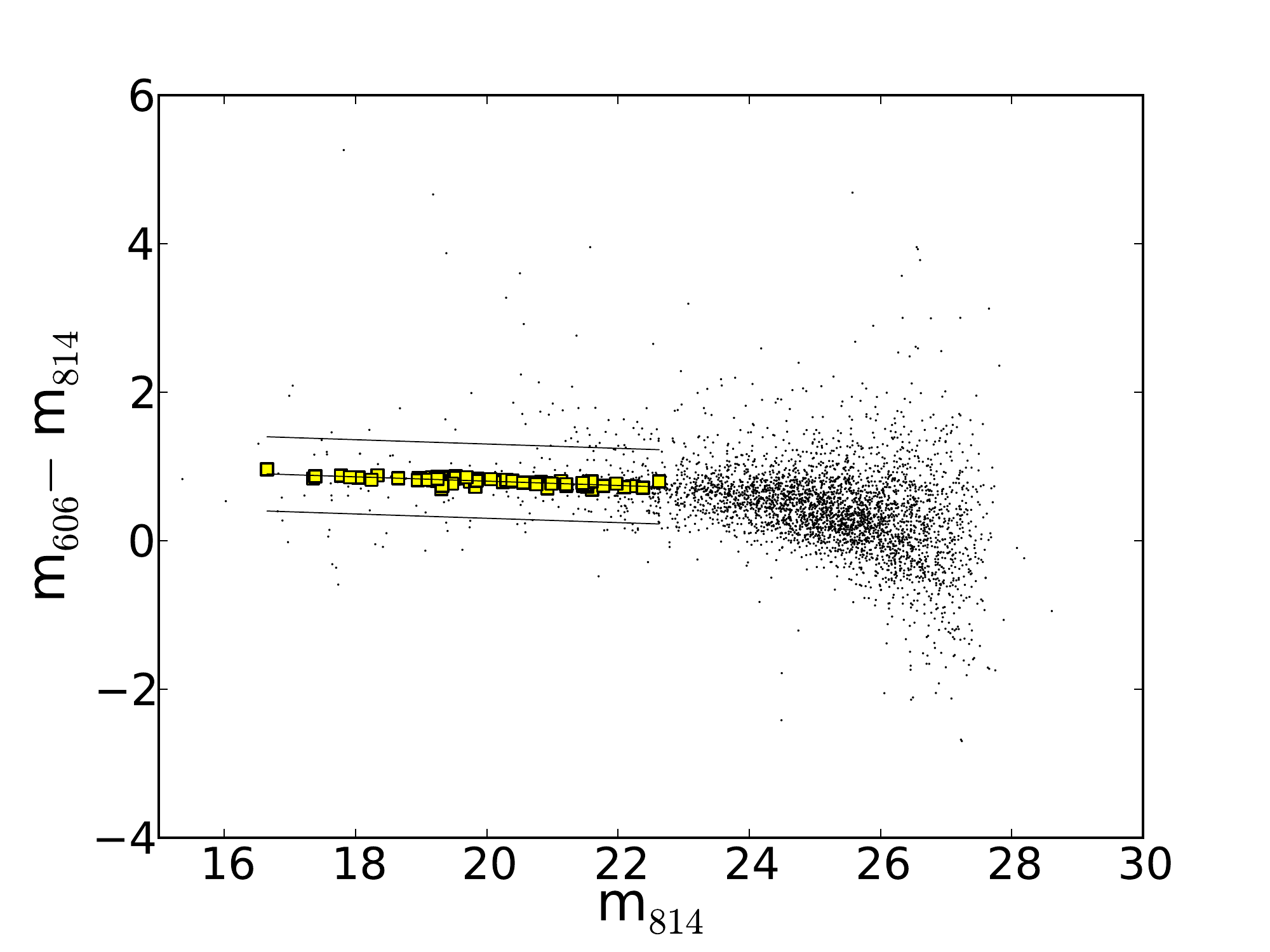}
	\caption{Perturbers to the lens model are plotted as yellow squares. On the vertical axis colour is computed from \sourceextractor{} \citep{sourceextractor1996} aperture magnitudes, $40$ pixel diameter. The horizontal axis is \sourceextractor{} auto magnitudes.  Magnitudes are computed from {\it HST} images. A trendline is plotted through the perturbers with additional lines above and below to show $\pm0.5$ shift in colour of the trendline.  All of the perturbers are on the red sequence for the cluster.}
	\label{figCM606814}
\end{figure}

The cluster members act only as perturbers in the lens model.  Their impact is more pronounced when near critical curves. Perturbers were selected from the red cluster sequence of a colour-magnitude diagram, as shown in Figure~\ref{figCM606814}, and confirmed with visual inspection of the {\it HST} images that the perturbers were galaxies and not  local stars or noise. The perturbers were required to lie on the red sequence in both of the colour-magnitude diagrams. A majority of these galaxies have spectroscopic redshifts that were used to establish cluster membership where available \citep{white2146dynamics}.

\subsection{Model Refinement} \label{SecModelRefinement}
Strongly lensed objects are used for constraining a parametric lens model. Of the morphology of a multiple image system, only the positions of the images in the lens plane are used in this paper.  The error assumed on the positions of images was $0.3$ arcsec to account for telescope resolution as well as substructure in the lens \citep{massey2015}.

\begin{figure}
	\includegraphics[width=\columnwidth]{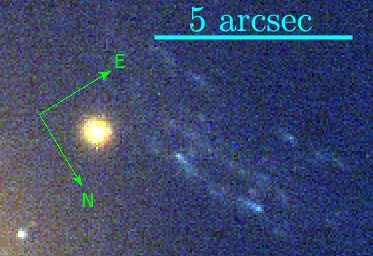}
	\caption{Close-up view of a false colour image of the `mask.'  This system is East of \bcga{}.  This feature exhibits a bilateral symmetry.}
	\label{figSamuraiMask}
\end{figure}

\begin{figure}
	\includegraphics[width=\columnwidth]{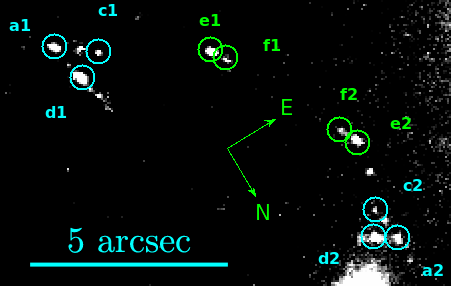}
	\includegraphics[width=\columnwidth]{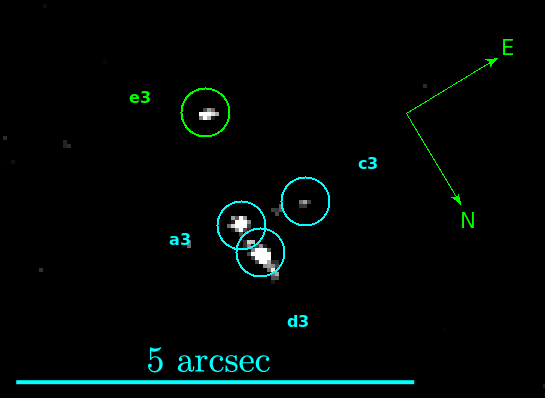}
	\caption{From the f606W filter, a close-up view of a high contrast image of the `bra-ket,' called so because the shape resembles `$<$' and `$>$.'  In the top panel the bilateral symmetry is evident and in the bottom panel there is a 3rd counter image.  The symmetric pair is located West of \bcga{} and the third counter image is North-East of \bcga{}. The `bra-ket' system is modelled with points $a$, $c$, and $d$.  The blue pair next to `$<$' and '$>$' is modelled with points $e$ and $f$.}
	\label{figBraket}
\end{figure}

The lens model was iteratively built, starting first with obvious multiple image systems, Figures~\ref{figSamuraiMask} and \ref{figBraket}.  A model was then used to predict and identify other multiply imaged objects. Section~\ref{MultipleImageDetection} describes this technique in detail. 


For extended, irregular background objects, such as an irregular galaxy or merging galaxies, concentrations of baryonic matter can appear as nodules.  The positions of the nodules were used as constraints in the lens model \citep[see for example][]{Sharon2014}.  For these nodules, the location of the brightest pixel was used as a constraint with an assumed error of $0.3$ arcsec or $6$ pixels in the {\it HST} images.  For isolated elliptical galaxies, the brightest pixel was used to identify the position of the object.  

For families of multiple images arising from the same source, the redshift of each of the images was constrained to be identical.

\subsubsection{Multiple Image Detection} \label{MultipleImageDetection}

\begin{figure*}
	\includegraphics[width=\textwidth]{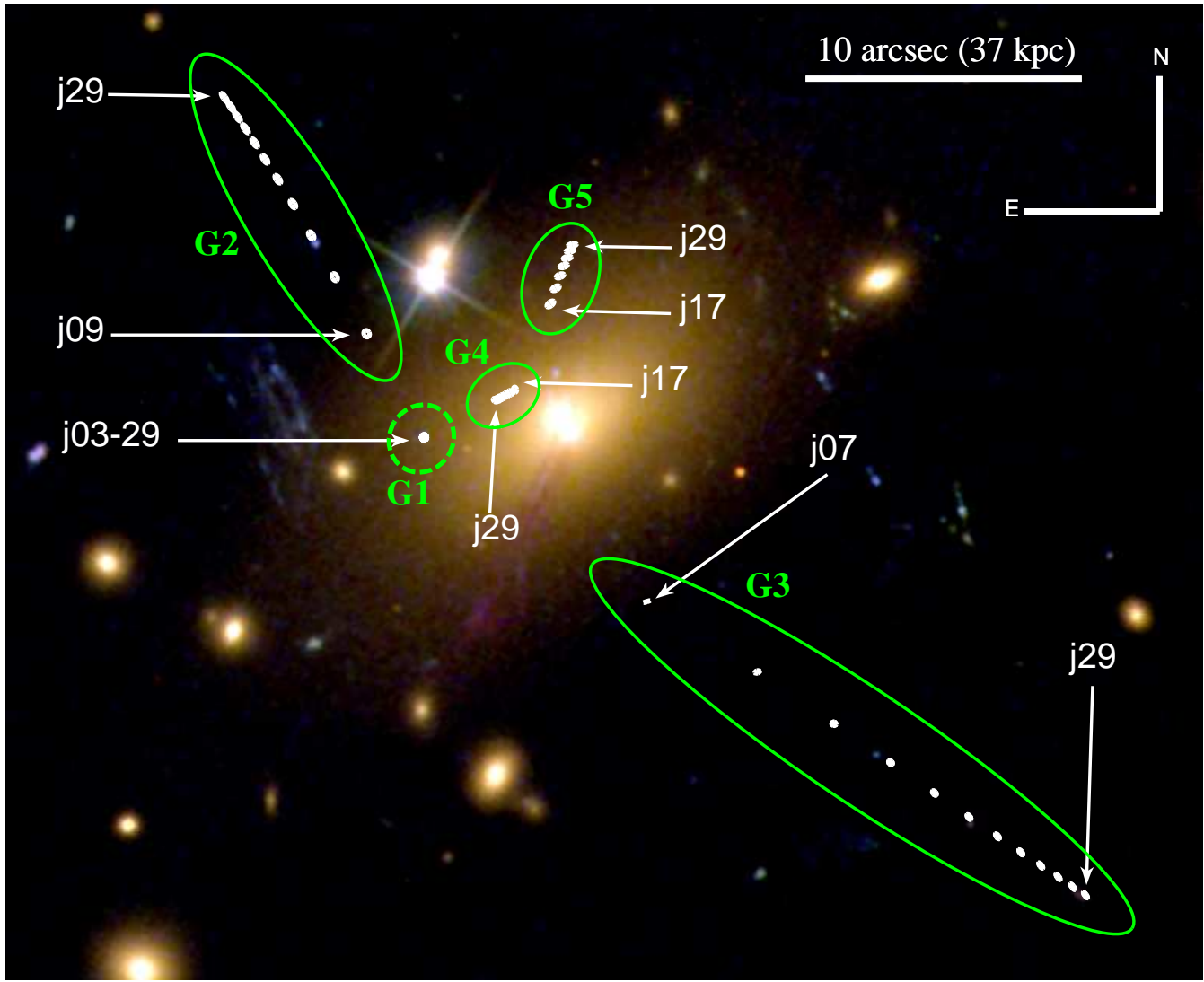}
	\caption{In the vicinity of \bcga{}, group G1 is the test point in the image plane.  Groups G2 through G5 are counter images of G1 parametrized by redshift.  The position of an image, G1, is projected from the image plane to the source plane at varying redshifts.  The resulting source plane positions of G1 at varying redshifts are then projected onto the lens plane. The multiplicity of an image is dependent on redshift.  This results in various predictions of multiple images in groups G2 through G5 as functions of assumed redshift of the source of the image in group G1. Group G2 begins at a redshift of $z=0.9$, group G3 begins at $z=0.7$, and groups G4 and G5 begin to appear at a redshift of $z=1.7$. }
	\label{figMulipleImageDetectionGroups}
\end{figure*}

The following iterative approach was used to predict multiple images.  Objects near a Brightest Cluster Galaxy (BCG) are considered to be within the Einstein radius and should be multiply imaged if far enough behind the lens. With a candidate image of a source identified, \lenstool{} was used to predict corresponding counter images for the source, for different source redshifts.  With a point in the image plane, a lens model was used to deproject that point onto the source plane at an array of different redshifts.  These deprojected points were then reprojected with the lens model onto the lens plane. Such parametrized-redshift counter images trace one or more lines in the image plane.  In Figure~\ref{figMulipleImageDetectionGroups} an object near \bcga{}, designated group G1, was tested.  The source redshift was parametrized in increments of 0.2 starting at $z = 0.3$ and ending at $z = 2.9$.  We limit the range for which we explore the redshift of strongly lensed objects to an upper bound of $z = 3$ because higher redshift objects will be fainter compared with the multiple image systems of interest, and will also drop out of the bluest filter.  The graduations in the parametrization of $z$ is arbitrary. The step size of $0.2$ used here is for illustrative purposes only.  Smaller step sizes were used, on the order of $0.1$.

As can be seen in Figure~\ref{figMulipleImageDetectionGroups}, the groups of predicted counter images form lines in the image plane. The mass model used to compute this was an intermediate iteration utilizing a  subset of the finalized image constraints.  The image multiplicity is dependent on the redshift of the lensed object since the caustic structure changes as a function of redshift. Group G2 contains counter images which first appear at source redshift $z = 0.9$.  There are no predicted counter images for source redshift $z = 0.3$, $0.5$, or $0.7$ in group G2. In group G3, there were counter images starting at source redshift $z = 0.7$. In groups G4 and G5, counter images started to appear at source redshift $z = 1.7$.

\begin{figure}
	\includegraphics[width=\columnwidth]{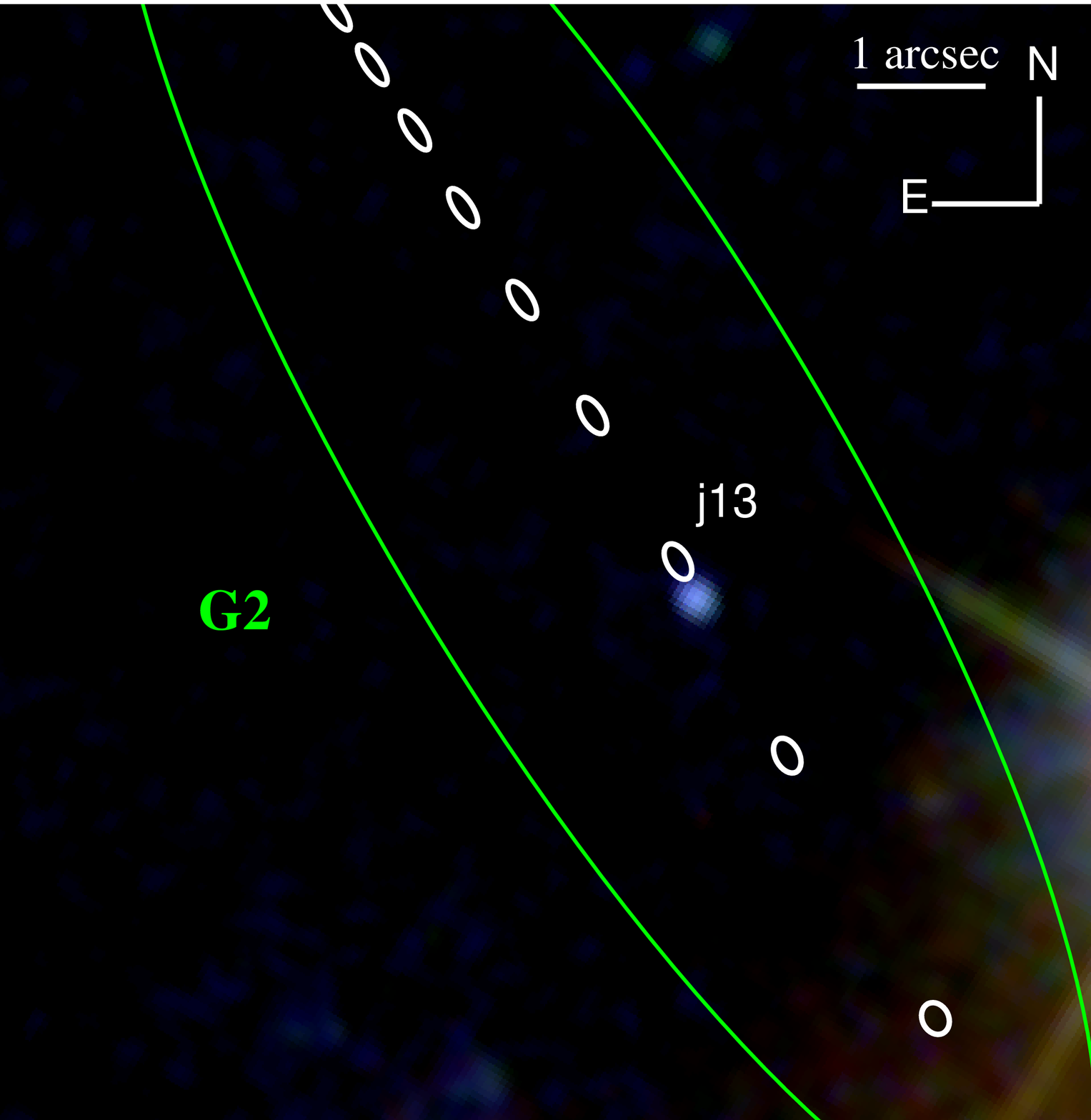}
	\includegraphics[width=\columnwidth]{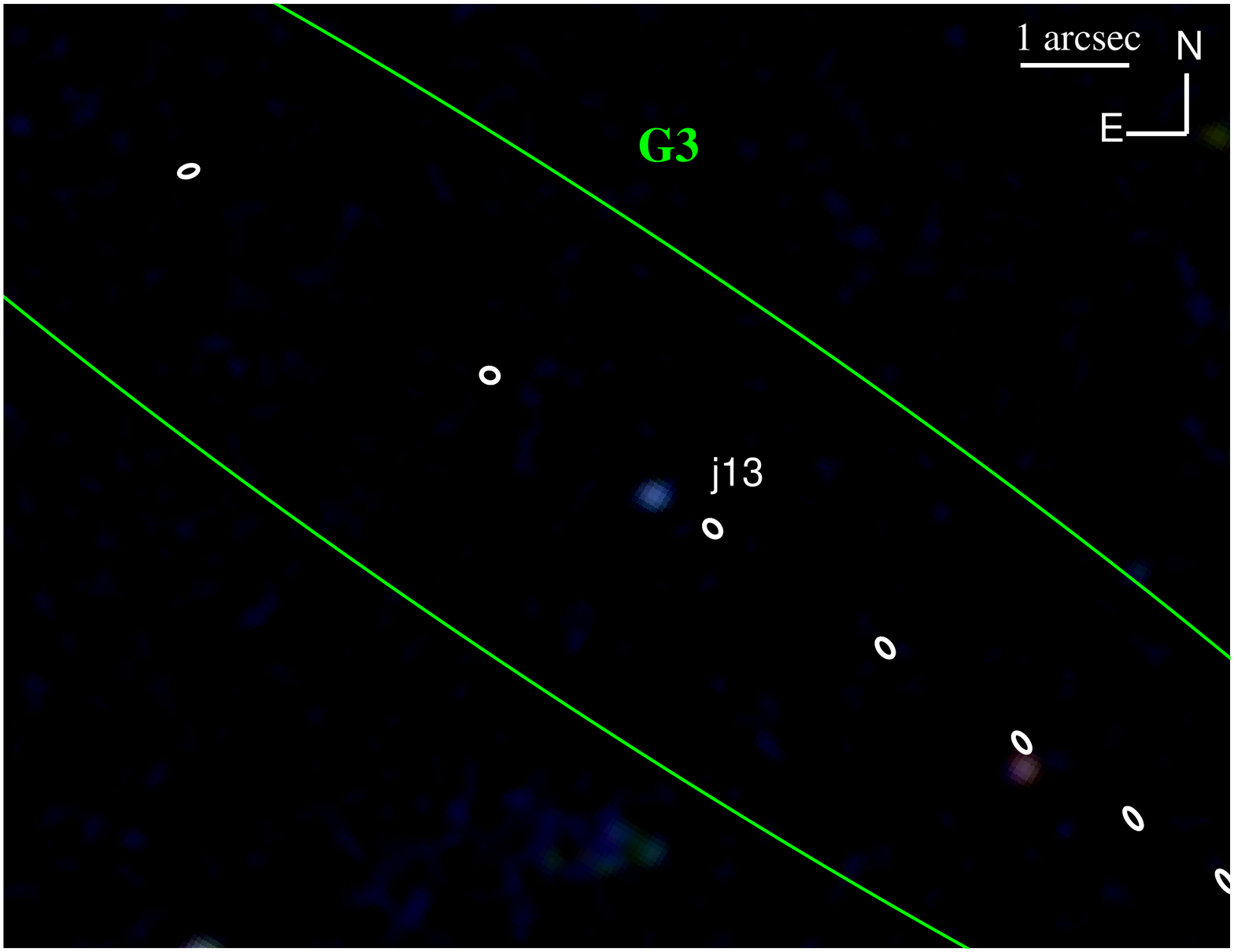}	
	\caption{In the top frame, an expanded view of the region around G2 in Figure~\ref{figMulipleImageDetectionGroups}.  There is a blue galaxy as a possible counter image for some assumed redshift of G1 slightly less than $Z = 1.3$ for some intermediate mass model.  The  counter images in this group have an increasing redshift in the North East direction.  In the bottom frame, a blue galaxy in group G3 appears as a possible counter image for the same assumed redshift of G1 between the same range of redshifts as the galaxy in the top frame.  Both of these galaxies are on the redshift parametrized path at the same $z$ location.  }
	\label{figMulipleImageDetectionGroups2}
\end{figure}

Objects near the predicted counter image line were considered candidates for a multiply imaged system.  Colour and morphology were taken into consideration.  For the lensed source depicted in Figure~\ref{figMulipleImageDetectionGroups}, images appeared on both of the lines in groups G2 and G3 near $z = 1.3$, see Figure~\ref{figMulipleImageDetectionGroups2}.   No objects were found on the lines defined by groups G4 and G5.  If the test object were to be a member of a multiple image system near redshift $z = 1.3$, then there should be no counter image in groups G4 and G5 since images of those groups do not occur for objects near redshift $z = 1.3$.   The three objects taken as a triplet are all blue and are of comparable brightness, and so was considered to be a multiply imaged source.  

Multiply imaged objects were added to the model as a further refinement.  This second procedure was iterated such that a refined model with extra constraints were used to make further predictions for additional possible multiple image systems.  Predicted multiple image systems that worsened the best fit of the lens model were discarded. 

\subsection{Lens Model Constraints} 
Multiple image systems were used as constraints and are presented here. In the refinement state of generating lens models, more systems were used than are presented here.  Specifically, the multiple image systems  and arcs around the second brightest cluster galaxy, \bcgb{}, were not used in the final models presented here, but were used in the intermediate stages for some variations to explore parameter space and to test convergence of the models.    
	
There are four main groups of multiple image systems near \bcga{}.  An object is denoted by a letter followed by a unique ordinal, where objects with the same first letter are considered to belong to the same multiple image system.  In Figure~\ref{figMultipeImages}, corresponding lensed images considered to be at the same redshift are denoted with the same colours.

\subsection{Computation}

For \lenstool{}, optimization was performed in the source plane in the refinement stage of the lens model construction where various multiple image systems were considered.  Final models were optimized in the image plane.  One thousand samples were taken for each of the image plane optimized models.  For optimization in the source plane, the $\chi^2$ fit uses the computed source positions relative to the barycenter of the multiple images in the source plane \citep{lenstool2007}.  The image plane optimization is more computationally expensive as the $\chi^2$ minimization must solve the lens equation to find counter images in the lens plane \citep{lenstool2007}.  The $\chi^2$ contribution for lens plane optimization, equation 9 in \citet{lenstool2007}, is for some source $i$ 
\begin{equation}
\chi^{2}_{\rm i} = \sum_{\rm j=1}^{n_{\rm i}} \frac{\left[ x^{\rm j}_{\rm obs} - x^{\rm j}(\bmath{\theta}) \right]^{2} }{\sigma^{2}_{\rm ij}},
\label{eqnChisquared}
\end{equation}
where $n_{\rm i}$ is the number of multiple images for $i$, $x^{\rm j}_{\rm obs}$ is the observed location, $x^{\rm j}(\bmath{\theta})$ is the position predicted by the model, and $\sigma^{2}_{\rm ij}$ is the error of the position of $j$, which is assumed to be $0.3$ arcseconds.  The $\chi^{2}$ value is used as a parameter in the likelihood function \citep{lenstool2007}.
 
The so-called Rate parameter, as defined in \citet[][see eq. 13]{lenstool2007}, was set to $0.1$.  This parameter controls the speed at which the modelling converges. The value used for the Rate is on the lower end of the range suggested in \citet{lenstool2007}.  Larger values for the final lens models did not alter the main results of this paper.

For \glafic{}, optimization was performed in the source plane but with
the full magnification tensor to convert offsets in the source plane
to those in the image plane. It has been shown that the resulting
$\chi^{2}$ computed in the source plane accurately reproduces $\chi^{2}$
directly computed in the image plane \citep{glafic_paper2010}. 

\section{Results and Discussion} \label{SecResults}

In this Section the results are presented for several image plane optimized models.  In all models the `bra-ket' system, composed of objects `a,' `c,' and `d,' shaped like `$<$' and `$>$', was statically assigned a redshift of $z = 2.0$.  This designation was used to break a degeneracy between the redshift of the multiply lensed objects and the mass of the lens.  We expect that images would drop out of the f435W filter near a redshift of $z = 3.0$.  The models explored a flat prior on source redshift for each of the multiple image systems over a range of redshift from $z = 0.3$, which is just beyond the cluster redshift, to $z = 3.0$. Other values for the redshift at which the bra-ket system was statically assigned were explored, and had no impact on the conclusions regarding the location of the Dark Matter Halo A.

The convergence $\kappa$, or dimensionless surface mass density \citep{Schneider1995book}, of the lens is a function of source redshift due to the dependence on the critical surface mass density $\Sigma_{\rm crit} = c^2 D_{\rm S}/(4\pi G D_{\rm L}D_{\rm LS})$ where $D_{\rm S}$, $D_{\rm L}$ and $D_{\rm LS}$ are the source, lens and lens-source angular diameter distances respectively.
In this paper we are concerned with the centroid of the mass, which is rather insensitive to the redshifts of the lensed sources adopted. We have established this by using different assigned redshifts in the lens modeling. Lack of knowledge of the source redshifts does have an impact on the normalisation of the mass profile, and the determination of the total mass. In any case the choice of $z = 2.0$ for a static redshift for the `bra-ket' multiple image system does not significantly alter the mass. The variation in mass was less than an order of magnitude.   For example, comparing the convergence for a source at redshift $z = 1.0$ and at $z = 2.0$ we obtain 
\begin{equation}
\kappa_{z=1} = \left( \frac{\Sigma_{{\rm crit},z=2}}{\Sigma_{{\rm crit},z=1}} \right) \kappa_{z=2} = \left( 0.86 \right)  \kappa_{z=2}, 
\label{eqnKappadegeneracy}
\end{equation}
which shows that the particular choice of redshift for one of the sources does not appreciably change the mass.

The choice of a redshift for a source image is to break a degeneracy for the mass of the halo and is arbitrarily chosen to be within the range of redshift of the cluster at $z=0.2323$ and $z=3.0$, where the upper bound of $z=3.0$ is where we expect dropouts to occur for given {\it HST} filters used in this study.  The degeneracy is evident from Equation~\ref{eqnKappadegeneracy} where, for some projected surface mass density $\Sigma$, we have $\Sigma = \kappa \Sigma_{\rm crit} $ \citep{Schneider1995book}.  With the redshift of the lens known the term $\Sigma_{\rm crit}$ is a function of the redshift of the source. From the geometry of the source images we determine $\kappa$.  However, $kappa$ is a ratio of the mass of the lens via the projected surface density $\Sigma$ and $\Sigma_{\rm crit}$, which are both unknown.  This gives a degeneracy between mass of the lens and the redshift of the sources.

For NFW profiles the concentrations were also held fixed.  We expected the turnover radius to be well beyond the field of view of the {\it HST} images.  With no information to constrain this radius since strong lensing features are typically only found in a relatively small area, the solution is simply hold the scale radius, $r_{s}$ to be fixed and allow the concentration to be a free parameter.  However, in this paper the concentration was fixed and the scale radius was a free parameter.  Both parameters are related by $c r_{s} = r_{200}$.   Holding $c$ fixed with $r_{s}$ free is the same as having $c$ free and $r_{s}$ fixed. 
 
We found that the position of the dark matter halo associated with Abell 2146-B was not well constrained, due to the presence of many massive galaxies close to the BCG contributing to strong lensing, and more importantly a lack of redshift information or clear corresponding morphological features for arcs to identify corresponding multiple images. 

The following are examples of candidate multiple image systems that were used in order to try to constrain the mass distribution of Abell 2146-B: (i) There is a large arc near \bcgb{} which was used as a constraint, however it is near to a group of 4 galaxies that are perturbing the shape of the arc. 
(ii) There is also a faint blue feature in the stellar halo of \bcgb{} that appears to be a multiple image system.  
(iii) Additionally, East of \bcgb{} there is an elongated arc that looks multiply imaged.  This object is likely to be relatively high redshift since it drops out of the f435W and f606W filters.  However, it is detected at a low signal to noise ratio in f814W, so the detection threshold in the other filters have to be accounted for in future studies of Abell 2146-B.

The refinement technique described in \S~\ref{SecModelRefinement} for predicting multiple image systems yielded unsatisfactory models.  An unsatisfactory model is one where the model fails to converge for one or more parameters.  With the model failing to converge on a solution it becomes problematic in sampling from the distribution to create a predictive intermediate model. This may also indicate that the dark matter is disrupted near \bcgb{} such that the matter distribution is not a smooth NFW profile due to the merger.  In early stages of model refinement, the position of the DM halo associated with \bcgb{} was allowed to vary.  However, since the location of DM halo B never converged to an acceptable solution due to lack of multiple image systems as constraints, in later stages of model refinement the DM halo B was set to a static position coincident with \bcgb{}, and models with two fixed haloes to describe Abell 2146-B were also considered.

Since we were not able to constrain the location of DM halo B, the final models presented below have static NFW profiles associated with Abell 2146-B, and the multiple image systems associated with Abell 2146-B are discussed no further.  The main science goal was to determine the location of the dark matter halo associated with Abell 2146-A.  The static haloes representing Abell 2146-B are far enough away from Abell 2146-A such that the only major effect is to provide some shear and very slightly offset the location of Halo A.  The shift in position is $1.2$ arcsec due to the additional fixed parameter haloes.  

We present three lens models here designated as model 1, 2 and 3.  In all models, there is a dark matter NFW halo associated with Abell 2146-A, Halo A.  The major distinctions between the models are specified below.

\textbf{Model 1} \label{secModelNaught}
	This model used only one NFW profile associated with Abell 2146-A.  The concentration was held fixed at $c = 3.5$.  The other parameters of the NFW profile were free as described earlier.  See Table~\ref{tblModelParameters} for a side by side comparison of fixed parameters between the models.  

\textbf{Model 2} \label{secModelBeta}
	In addition to the NFW halo associated with Abell 2146-A in Model 1, this model had a second dark matter halo placed at the location of \bcgb{}.  This second halo was static in the sense that all parameters, namely $M_{200}$, $c$, position, ellipticity, and position angle were fixed.  This gives this model a total of two NFW profiles to describe the cluster.
	
	The placement of the second NFW profile assumed that light traces matter.  The ellipticity and position angle of \bcgb{} were used to describe the dark matter halo, in addition to the location of the DM halo being coincident with the peak of \bcgb{}.

    Mass and concentration parameters of the NFW profile  obtained from earlier models were used as a reference for these statically assigned parameters.  The mass and concentration was chosen to be consistent with the weak lensing analysis by \citet{king2146weak} such that they were of the same order.  The weak lensing analysis by \citet{king2146weak} looked at concentrations between $3.5$ and $4.5.$, however intermediate strong lensing models resulted in lower concentrations at around $c=3.1.$  The mass of Halo B from intermediate models was on the order of $10^{14}\solMass$ where a specific choice of $4\times10^{14}\solMass$ obtained from an intermediate strong lens model was used in the final models.  The scale radius was determined from the assumed mass and concentration.
    
\textbf{Model 3} \label{secModelAlpha}
	This model has a total of three NFW profiles. 
	The dark matter associated with Abell 2146-B was represented with two NFW profiles.  One of the profiles was centred and fixed on the \bcgb{}, Halo B, as was done in Model 2, and the other was fixed at the location of the centroid of a mass peak seen in the weak lensing analysis by \citet{king2146weak}, Halo C.  See Figure~\ref{figA2146} for the location of Halo C.  While the presence of a mass peak at the location of Halo C is indicated by the weak lensing analysis \citep{king2146weak}, there are no obvious strongly lensed features associated with this peak that might allow us to better constrain its properties.

	The observed ellipticity and position angle of the \bcgb{} were used to describe the NFW halo B centred at the same location, i.e. light traces mass.  Halo C was assumed to be spherical, since there are no strong constraints from weak or strong lensing.  

	The total mass of Halo B from model 2 was divided equally between Halo B and Halo C of model 3.  

\textbf{Other Models} \label{secModelGamma}
	
	Additional models were considered with \bcgb{}  having free parameters constrained by multiple image systems nearby. However, as noted above, these models were unable to converge to a stable solution. Therefore, associated arcs and multiple images near \bcgb{} were not used for any of models 1, 2, or 3 above.


\begin{table}
	\caption{Comparison of distinguishing features between lens models 1, 2, and 3.  Mass is given in units of $10^{15} \solMass$. All parameters of Halo B and Halo C are fixed. }
	\label{tblModelParameters}
	\begin{tabular}{@{}lcccc}
	\hline
		& 			 & Model 1 & Model 2 & Model 3 	 \\
	\hline
	\multicolumn{5}{l}{Halo A} \\
		& c			 & 3.5 	& 3.5 		 & 3.5 \\
	\multicolumn{5}{l}{Halo B} \\
		& $\epsilon$ & -		& 0.177 	 & 0.177  	\\
		& PA		 & -		& 141.9		 & 141.9	\\
		& M$_{200}$	 & -		& 0.4148     & 0.2074   \\	
		& c			 & -		& 3.1		 & 3.1       \\
	\multicolumn{5}{l}{Halo C} \\
		& $\epsilon$ & -		& -			 & 0		\\
		& PA		 & -		& -			 & 0		\\
		& M$_{200}$	 & -		& -		     & 0.2074	\\
		& c			 & -		& -			 &  3.1		 \\
	\hline
	\end{tabular}
\end{table}

The associated gas of the cluster was considered as a perturbation in intermediate versions of the lens model in order to establish its significance in the modeling.  The impact of the gas was to displace the centroid of the DM Halo A away from the gas peak, towards \bcgb{}.  In the final models presented here, we do not include a gas component in the model.   The results here are therefore a lower bound on the separation between the dark matter halo and the X-ray cool core, since a gas component at the location of the cool X-ray core only increased the separation.

In Table~\ref{tblModelResults} Model 1, which had only one dark matter NFW halo, had a better $\chi^2 / \text{dof}$ compared to Model 2 and 3.  However, we do expect there to be dark matter near \bcgb{} as evidenced by the weak lensing analysis \citep{king2146weak}.  Model 2 had the largest Ln(Evidence) \citep{BayesInTheSky2008Trotta} value relative to Models 1 and 3 which indicates a preference for two dark matter haloes when using Ln(Evidence) to classify goodness of fit of models.  By considering the Bayes factor and Jeffrey's scale from \citet{BayesInTheSky2008Trotta}, which we can compute by exponentiating the difference of the Ln(Evidence), we can see that Model 3 is weakly preferential to Model 1, and that Model 2 is moderately preferential to either of the other two models.

\begin{figure}
	\includegraphics[width=84mm]{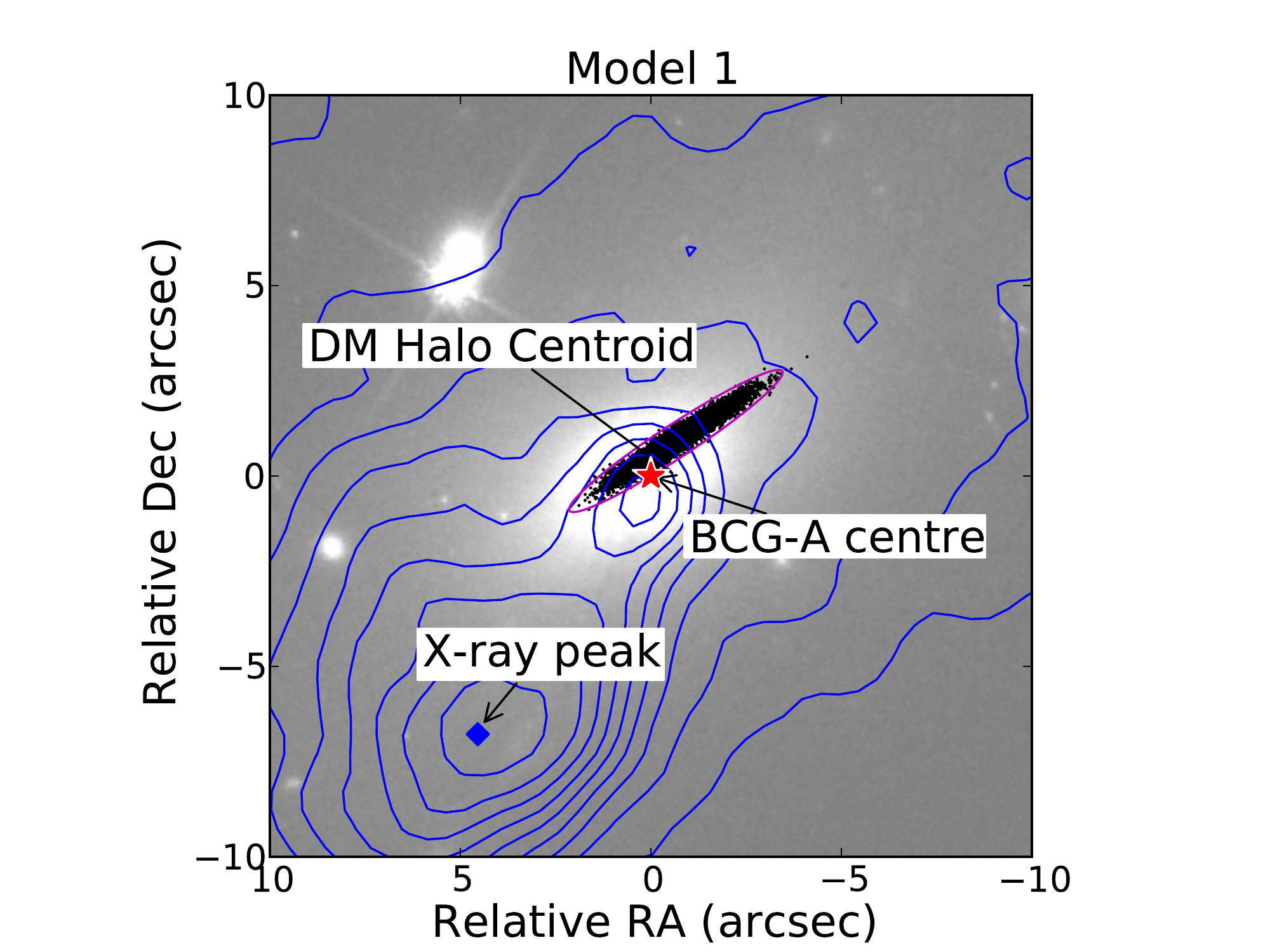}
	\includegraphics[width=84mm]{figure_555_6.pdf}
	\includegraphics[width=84mm]{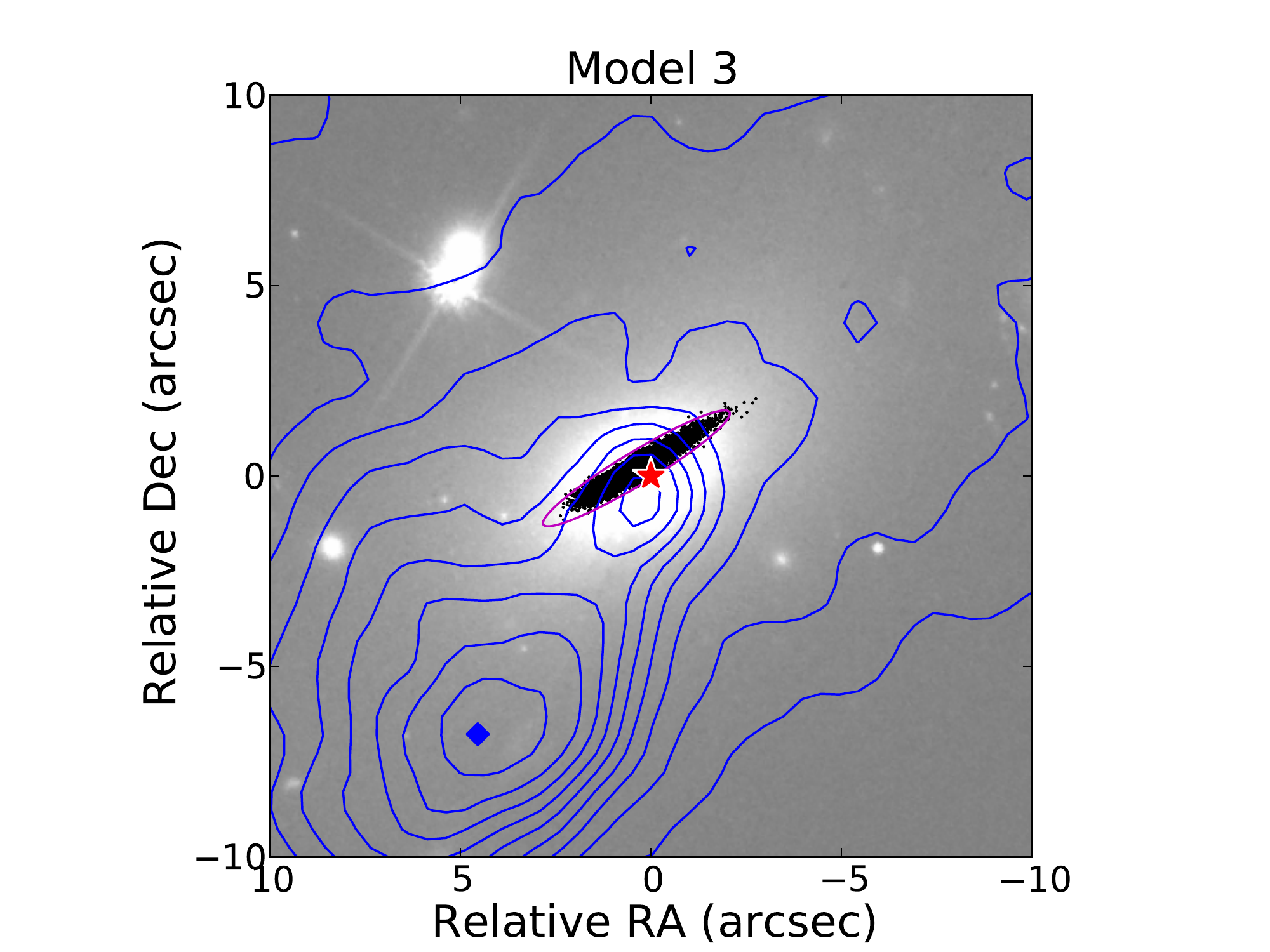}
	\caption{The red star is the centre of the \bcga{} as determined with \sourceextractor{} and the blue diamond is the peak of the hot X-ray gas \citep{russel2010abel2146, russellxray1}.  The collection of black points near the \bcga{} centre is a plot of the centroid of dark matter halo for 1000 samples from the best fit lens model.  North is up and East is left.  Top image is model 1, middle image is model 2, and bottom image is model 3.  \bcgb{} is out of the frame to the North West (or top right) direction from \bcga{}.  The contours are from {\it Chandra X-ray Observatory} with a 3 pixel Gaussian smoothing. Note that the second X-ray peak coincident with \bcga{} is a point source associated with the AGN of the BCG.  This is not a gas peak.  The X-rays from this point source are non-thermal  \citep{russellxray1}.}
	\label{figResults518}
\end{figure}

All models considered, including versions not presented here, resulted in a well constrained centroid for dark matter halo A.  In each model, the DM halo A was coincident with \bcga{}, and thus lagging behind the leading gas peak.  In Figure~\ref{figResults518} there are $1000$ samples from the best model obtained from \lenstool{} for Models 1, 2, and 3 which are plotted over the HST image.  A $3\sigma$ curve is drawn around the samples.  All of the final models presented here result in the \bcga{} peak being around approximately $3\sigma$ of the mean location of the Dark Matter Halo A.   The presence of mass associated with Dark Matter Halo B or C shift the centroid of Halo A away and towards the X-ray peak.

As a consistency check, the strong lensing program \glafic{} \citep{glafic_paper2010} was used to verify model selection.  In some instances of models that don't converge to a solution, or when the specified error on image positions, $\sigma_{\rm ij}$ in Equation~\ref{eqnChisquared}, is too small,  \lenstool{} will fail to generate a model that can reproduce the input constraints.  This is most likely due to source positions being very close to a complicated caustic line. 
As a measure of the quality of a lens model we expect that solutions for source positions corresponding to image position constraints are properly projected from their corresponding source plane position to the lens plane position which was used as a constraint in constructing the model.  In other words, for a given image used to construct a lens model that predicts a source position, we expect to use the lens model and project the source position and recover the image positions used in constructing the model.  For \lenstool{}, this can occur in the cases mentioned above.  \glafic{} allows for requiring solutions to match the number of aforementioned projected images with the number of images used as constraints in building the lens model.  Both of these behaviours are complementary in constructing lens models.

The error and location of DM halo A was consistent with results produced with \lenstool{}.  For \glafic{} Models 1, 2, and 3 the separation distances between the dark matter halo centroid and \bcga{} are, respectively, approximately $5$, $3$, and $6$ {\rm kpc}.   The \glafic{} Models 1, 2, and 3 separation distances between the dark matter halo and the X-ray peak are, respectively, $28$, $29$, and $26$ {\rm kpc}.  Each respective result from the three \glafic{} models for the DM Halo A position was within $3\sigma$ for all three \lenstool{} models. That is to say that DM Halo A position results from \glafic Model 1 were within the $3\sigma$ boundary for each of the \lenstool{} models, and so forth for each \glafic{} model.

\begin{table}
	\caption{Comparison of results from lens models 1, 2, and 3. The Mean Position is for the dark mater centroid.  The error matrix shows the variance and covariance of the dark mater halo position. The separations are between the mean position of the dark matter centroid and either the peak light position of \bcga{} or the X-ray peak.  The Ln(Evidence) for Model 2 yields a significant Bayes factor \citep{BayesInTheSky2008Trotta} relative to either of the other models.}
	\label{tblModelResults}
	\begin{tabular}{@{}lccc}
		\hline
					 & Model 1 	& Model 2 & Model 3	\\
		\hline 	
		\multicolumn{4}{l}{Mean Position (deg)} \\
		RA							& $239\fdg05789$	& $239\fdg05860$ & $239\fdg05861$ \\
		Dec							&  $66\fdg348371$  & $66\fdg348183$ & $66\fdg348171$ \\
		\hline
		 \multicolumn{4}{l}{Error Matrix (arcsec)}\\
		 $\sigma^{2}_{\rmn{RA}}$	 & 0.7 & 0.5 	& 0.5 	 \\
		 $\sigma^{2}_{\rmn{Dec}}$	 & 0.3 & 0.2 	& 0.2 	 \\
		 $\rmn{Cov(RA,Dec)}$		 & 0.4 & 0.3 	& 0.3 	 \\
		 \hline
		 \multicolumn{4}{l}{Separation Distances (kpc)} \\
		 \bcga{}			         & 4.19 &  1.63  	& 1.59 	 \\
		 X-ray peak		             & 34.5 &  30.3	    & 30.2	 \\
		\hline
		Ln(Evidence)	             & 36.8  & 41.1	& 37.8   \\
		$\chi^{2} / \text{dof}$      & 0.51  & 0.38 & 0.18 \\
		\hline
	\end{tabular}
\end{table}

\section{Conclusion} \label{secConclusion}

We have used {\it Hubble Space Telescope} images to identify strongly lensed multiple image systems near to the BCGs of the merging cluster, Abell 2146.  
This merging cluster is an important laboratory for studying the physics of cluster mergers because the collision has occurred near the plane of the sky, the two 
clusters in the system are of comparable mass and the system also has two well defined shock fronts. The BCG in Abell 2146-A has an X-ray cool core.  
By identifying multiple image systems in the centre of Abell 2146-A, we have made a strong gravitational lensing mass model. We have determined that the 
location of the dark matter halo is coincident with the BCG, and is offset from the X-ray cool core. In other words from this strong lensing analysis, the cool 
core is leading, rather than lagging, the dark matter post collision, contrary to expectations for a merger seen shortly after first core passage.

In \citet{canning2012_a2146}  it is proposed that there is a causal link between the X-ray cool core and a plume of gas extending from the BCG in its direction, which is spatially coincident with soft X-ray emission. Together with the disrupted nuclear 
structure of the BCG, an interaction between the BCG and another galaxy in the cluster (prior to or during the merger) is proposed as a possible explanation 
of the offset between the BCG and the X-ray cool core \citep{canning2012_a2146}.

\citet{Hamer2012MNRASthreeclustersgascooling} describe three clusters with significant offsets between the BCG and X-ray peak with the conclusion that a transitory event is the source of decoupling.  The largest separation distance in \citet{Hamer2012MNRASthreeclustersgascooling} is on the order of $10$\,kpc, however, whereas in Abell 2146 the offset is on the order of $30$\,kpc.   

However, in light of \citet{Sanderson2009}, Abell 2146-A seems to fall in the upper range of offsets for BCG with line emission \citep{Crawford1999}.

In order to test these hypotheses and to better understand the distribution of the dark matter (and galaxies), and plasma in the aftermath of the collision, 
computer simulations are being undertaken. These simulations will be guided by the results of lensing, X-ray, SZ and additional data.

Redshift information for the strongly lensed features is critical to obtaining a properly normalized strong lensing mass model for Abell 2146-A. This redshift information would in addition allow us to identify multiple image systems to use as constraints in mapping the distribution of mass in Abell 2146-B. For objects that are bright enough, spectroscopic redshifts would be ideal, and otherwise photometric redshifts can be obtained with the addition of near-infrared imaging data to complement our optical imaging data. More work needs to be done to constrain the dark matter near \bcgb{}.  A combined weak and strong lensing analysis will help accomplish this.

\section*{Acknowledgements}
JEC acknowledges support from The University of Texas at Dallas, and NASA through a Fellowship of the Texas Space Grant Consortium.  

Based on observations made with the NASA/ESA Hubble Space Telescope, obtained through program 12871 through the Space Telescope Science Institute, which is operated by the Association of Universities for Research in Astronomy, Inc., under NASA contract NAS 5-26555.  Additional funding supporting JEC, LJK, and DIC came from a grant from the Space Telescope Science Institute under the same program 12871.  Additional funding supporting JEC and LJK came from a grant from the National Science Foundation, number 1517954.

This work was supported in part by World Premier International Research Center Initiative (WPI Initiative), MEXT, Japan, and JSPS KAKENHI Grant Number 26800093 and 15H05892.

We would like to graciously thank Keren Sharon for valuable comments and help.

We would like to thank the anonymous referee for very helpful comments and questions.  This work has significantly benefited from the efforts of the referee.

\bibliography{abell2146_SL}{}

\begin{thebibliography}{}

\bibitem[\protect\citeauthoryear{{Bertin} \& {Arnouts}}{{Bertin} \&
  {Arnouts}}{1996}]{sourceextractor1996}
{Bertin} E.,  {Arnouts} S.,  1996, A\&AS, 117, 393

\bibitem[\protect\citeauthoryear{{Canning}, {Russell}, {Hatch}, {Fabian},
  {Zabludoff}, {Crawford}, {King}, {McNamara}, {Okamoto} \&
  {Raimundo}}{{Canning} et~al.}{2012}]{canning2012_a2146}
{Canning} R.~E.~A.,  {Russell} H.~R.,  {Hatch} N.~A.,  {Fabian} A.~C.,
  {Zabludoff} A.~I.,  {Crawford} C.~S.,  {King} L.~J.,  {McNamara} B.~R.,
  {Okamoto} S.,    {Raimundo} S.~I.,  2012, MNRAS, 420, 2956

\bibitem[\protect\citeauthoryear{{Clowe}, {Brada{\v c}}, {Gonzalez},
  {Markevitch}, {Randall}, {Jones} \& {Zaritsky}}{{Clowe}
  et~al.}{2006}]{bullet3}
{Clowe} D.,  {Brada{\v c}} M.,  {Gonzalez} A.~H.,  {Markevitch} M.,  {Randall}
  S.~W.,  {Jones} C.,    {Zaritsky} D.,  2006, ApJ, 648, L109

\bibitem[\protect\citeauthoryear{{Crawford}, {Allen}, {Ebeling}, {Edge} \&
  {Fabian}}{{Crawford} et~al.}{1999}]{Crawford1999}
{Crawford} C.~S.,  {Allen} S.~W.,  {Ebeling} H.,  {Edge} A.~C.,    {Fabian}
  A.~C.,  1999, MNRAS, 306, 857

\bibitem[\protect\citeauthoryear{{El{\'{\i}}asd{\'o}ttir}, {Limousin},
  {Richard}, {Hjorth}, {Kneib}, {Natarajan}, {Pedersen}, {Jullo} \&
  {Paraficz}}{{El{\'{\i}}asd{\'o}ttir} et~al.}{2007}]{dPIE2007}
{El{\'{\i}}asd{\'o}ttir} {\'A}.,  {Limousin} M.,  {Richard} J.,  {Hjorth} J.,
  {Kneib} J.-P.,  {Natarajan} P.,  {Pedersen} K.,  {Jullo} E.,    {Paraficz}
  D.,  2007, {Where is the matter in the Merging Cluster Abell 2218?},
  Unpublished

\bibitem[\protect\citeauthoryear{{Hallman} \& {Markevitch}}{{Hallman} \&
  {Markevitch}}{2004}]{hallman2004}
{Hallman} E.~J.,  {Markevitch} M.,  2004, ApJL, 610, L81

\bibitem[\protect\citeauthoryear{{Hamer}, {Edge}, {Swinbank}, {Wilman},
  {Russell}, {Fabian}, {Sanders} \& {Salom{\'e}}}{{Hamer}
  et~al.}{2012}]{Hamer2012MNRASthreeclustersgascooling}
{Hamer} S.~L.,  {Edge} A.~C.,  {Swinbank} A.~M.,  {Wilman} R.~J.,  {Russell}
  H.~R.,  {Fabian} A.~C.,  {Sanders} J.~S.,    {Salom{\'e}} P.,  2012, MNRAS,
  421, 3409

\bibitem[\protect\citeauthoryear{{Harvey}, {Massey}, {Kitching}, {Taylor} \&
  {Tittley}}{{Harvey} et~al.}{2015}]{Harvey2015darkmattercrosssection}
{Harvey} D.,  {Massey} R.,  {Kitching} T.,  {Taylor} A.,    {Tittley} E.,
  2015, Science, 347, 1462

\bibitem[\protect\citeauthoryear{{Jaffe}}{{Jaffe}}{1983}]{jaffe1983}
{Jaffe} W.,  1983, MNRAS, 202, 995

\bibitem[\protect\citeauthoryear{{Jullo}, {Kneib}, {Limousin},
  {El{\'{\i}}asd{\'o}ttir}, {Marshall} \& {Verdugo}}{{Jullo}
  et~al.}{2007}]{lenstool2007}
{Jullo} E.,  {Kneib} J.-P.,  {Limousin} M.,  {El{\'{\i}}asd{\'o}ttir} {\'A}.,
  {Marshall} P.~J.,    {Verdugo} T.,  2007, New Journal of Physics, 9, 447

\bibitem[\protect\citeauthoryear{{Keeton}}{{Keeton}}{2001}]{keeton2001astro0102341}
{Keeton} C.~R.,  2001, ArXiv Astrophysics e-prints

\bibitem[\protect\citeauthoryear{{King}, {Clowe}, {Coleman}, {Russell},
  {Santana}, {White}, {Canning}, {Deering}, {Fabian}, {Lee}, {Li} \&
  {McNamara}}{{King} et~al.}{2016}]{king2146weak}
{King} L.~J.,  {Clowe} D.~I.,  {Coleman} J.~E.,  {Russell} H.~R.,  {Santana}
  R.,  {White} J.~A.,  {Canning} R.~E.~A.,  {Deering} N.~J.,  {Fabian} A.~C.,
  {Lee} B.~E.,  {Li} B.,    {McNamara} B.~R.,  2016, MNRAS

\bibitem[\protect\citeauthoryear{{Kneib}, {Ellis}, {Smail}, {Couch} \&
  {Sharples}}{{Kneib} et~al.}{1996}]{lenstool1996}
{Kneib} J.-P.,  {Ellis} R.~S.,  {Smail} I.,  {Couch} W.~J.,    {Sharples}
  R.~M.,  1996, ApJ, 471, 643

\bibitem[\protect\citeauthoryear{{Limousin}, {Richard}, {Jullo}, {Kneib},
  {Fort}, {Soucail}, {El{\'{\i}}asd{\'o}ttir}, {Natarajan}, {Ellis}, {Smail},
  {Czoske}, {Smith}, {Hudelot}, {Bardeau}, {Ebeling}, {Egami} \&
  {Knudsen}}{{Limousin} et~al.}{2007}]{Limousin2007}
{Limousin} M.,  {Richard} J.,  {Jullo} E.,  {Kneib} J.-P.,  {Fort} B.,
  {Soucail} G.,  {El{\'{\i}}asd{\'o}ttir} {\'A}.,  {Natarajan} P.,  {Ellis}
  R.~S.,  {Smail} I.,  {Czoske} O.,  {Smith} G.~P.,  {Hudelot} P.,  {Bardeau}
  S.,  {Ebeling} H.,  {Egami} E.,    {Knudsen} K.~K.,  2007, ApJ, 668, 643

\bibitem[\protect\citeauthoryear{{Markevitch}, {Gonzalez}, {Clowe},
  {Vikhlinin}, {Forman}, {Jones}, {Murray} \& {Tucker}}{{Markevitch}
  et~al.}{2004}]{bullet4}
{Markevitch} M.,  {Gonzalez} A.~H.,  {Clowe} D.,  {Vikhlinin} A.,  {Forman} W.,
   {Jones} C.,  {Murray} S.,    {Tucker} W.,  2004, ApJ, 606, 819

\bibitem[\protect\citeauthoryear{{Markevitch} \& {Vikhlinin}}{{Markevitch} \&
  {Vikhlinin}}{2007}]{markevitch2007}
{Markevitch} M.,  {Vikhlinin} A.,  2007, Phys Rep, 443, 1

\bibitem[\protect\citeauthoryear{{Massey}, {Williams}, {Smit}, {Swinbank},
  {Kitching}, {Harvey}, {Jauzac}, {Israel}, {Clowe}, {Edge}, {Hilton}, {Jullo},
  {Leonard}, {Liesenborgs}, {Merten}, {Mohammed}, {Nagai}, {Richard},
  {Robertson}, {Saha}, {Santana}, {Stott}, {Tittley}}{{Massey} et~al.}{2015}]{massey2015}
{Massey} R.,  {Williams} L.,  {Smit} R.,  {Swinbank} M.,  {Kitching} T.~D.,
  {Harvey} D.,  {Jauzac} M.,  {Israel} H.,  {Clowe} D.,  {Edge} A.,  {Hilton}
  M.,  {Jullo} E.,  {Leonard} A.,  {Liesenborgs} J.,  {Merten} J.,  {Mohammed}
  I.,  {Nagai} D.,  {Richard} J.,  {Robertson} A.,  {Saha} P.,  {Santana} R.,
  {Stott} J.,    {Tittley} E.,  2015, MNRAS, 449, 3393

\bibitem[\protect\citeauthoryear{{Meylan}, {Jetzer}, {North}, {Schneider},
  {Kochanek} \& {Wambsganss}}{{Meylan} et~al.}{2006}]{SchneiderSaasFee2006}
{Meylan} G.,  {Jetzer} P.,  {North} P.,  {Schneider} P.,  {Kochanek} C.~S.,
  {Wambsganss} J.,  eds, 2006, {Gravitational Lensing: Strong, Weak and Micro}

\bibitem[\protect\citeauthoryear{{Navarro}, {Frenk} \& {White}}{{Navarro}
  et~al.}{1996}]{NFW1996}
{Navarro} J.~F.,  {Frenk} C.~S.,    {White} S.~D.~M.,  1996, ApJ, 462, 563

\bibitem[\protect\citeauthoryear{{Oguri}}{{Oguri}}{2010}]{glafic_paper2010}
{Oguri} M.,  2010, PASJ, 62, 1017

\bibitem[\protect\citeauthoryear{{Postman}, {Coe}, {Ben{\'{\i}}tez}, {Bradley},
  {Broadhurst}, {Donahue}, {Ford}, {Graur}, {Graves}, {Jouvel}, {Koekemoer},
  {Lemze}, {Medezinski}, {Molino}, {Moustakas}, {Ogaz}, {Riess}, {Rodney},
  {Rosati}, {Umetsu}, {Zheng}, {Zitrin}, {Bartelmann}, {Bouwens}, {Czakon},
  {Golwala},  {Host}, {Infante}, {Jha}, {Jimenez-Teja}, {Kelson},
  {Lahav}, {Lazkoz}, {Maoz}, {McCully}, {Melchior},
  {Meneghetti}, {Merten}, {Moustakas}, {Nonino}, {Patel},
  {Reg{\"o}s}, {Sayers}, {Seitz}, {Van der Wel}}{{Postman} et~al.}{2012}]{Postman2012CLASH}
{Postman} M.,  {Coe} D.,  {Ben{\'{\i}}tez} N.,  {Bradley} L.,  {Broadhurst} T.,
   {Donahue} M.,  {Ford} H.,  {Graur} O.,  {Graves} G.,  {Jouvel} S.,
  {Koekemoer} A.,  {Lemze} D.,  {Medezinski} E.,  {Molino} A.,  {Moustakas} L.,
   {Ogaz} S.,  {Riess} A.,  {Rodney} S.,  {Rosati} P.,  {Umetsu} K.,  {Zheng}
  W.,  {Zitrin} A.,  {Bartelmann} M.,  {Bouwens} R.,  {Czakon} N.,  {Golwala}
  S.,  {Host} O.,  {Infante} L.,  {Jha} S.,  {Jimenez-Teja} Y.,  {Kelson} D.,
  {Lahav} O.,  {Lazkoz} R.,  {Maoz} D.,  {McCully} C.,  {Melchior} P.,
  {Meneghetti} M.,  {Merten} J.,  {Moustakas} J.,  {Nonino} M.,  {Patel} B.,
  {Reg{\"o}s} E.,  {Sayers} J.,  {Seitz} S.,    {Van der Wel} A.,  2012, ApJS,
  199, 25

\bibitem[\protect\citeauthoryear{{Press}, {Teukolsky} S.~A.~and{Vetterling} \&
  {Flannery}}{{Press} et~al.}{1992}]{NumericalRecipies}
{Press} W.~H.,  {Teukolsky} S.~A.~and{Vetterling} W.~T.,    {Flannery} B.~P.,
  1992, Numerical Recipes in C.
Cambridge University Press

\bibitem[\protect\citeauthoryear{{Randall}, {Markevitch}, {Clowe}, {Gonzalez}
  \& {Brada{\v c}}}{{Randall} et~al.}{2008}]{randall2008}
{Randall} S.~W.,  {Markevitch} M.,  {Clowe} D.,  {Gonzalez} A.~H.,    {Brada{\v
  c}} M.,  2008, ApJ, 679, 1173

\bibitem[\protect\citeauthoryear{{Russell}, {McNamara}, {Sanders}, {Fabian},
  {Nulsen}, {Canning}, {Baum}, {Donahue}, {Edge}, {King} \& {O'Dea}}{{Russell}
  et~al.}{2012}]{russellxray1}
{Russell} H.~R.,  {McNamara} B.~R.,  {Sanders} J.~S.,  {Fabian} A.~C.,
  {Nulsen} P.~E.~J.,  {Canning} R.~E.~A.,  {Baum} S.~A.,  {Donahue} M.,  {Edge}
  A.~C.,  {King} L.~J.,    {O'Dea} C.~P.,  2012, MNRAS, 423, 236

\bibitem[\protect\citeauthoryear{{Russell}, {Sanders}, {Fabian}, {Baum},
  {Donahue}, {Edge}, {McNamara} \& {O'Dea}}{{Russell}
  et~al.}{2010}]{russel2010abel2146}
{Russell} H.~R.,  {Sanders} J.~S.,  {Fabian} A.~C.,  {Baum} S.~A.,  {Donahue}
  M.,  {Edge} A.~C.,  {McNamara} B.~R.,    {O'Dea} C.~P.,  2010, MNRAS, 406,
  1721

\bibitem[\protect\citeauthoryear{{Sanderson}, {Edge} \& {Smith}}{{Sanderson}
  et~al.}{2009}]{Sanderson2009}
{Sanderson} A.~J.~R.,  {Edge} A.~C.,    {Smith} G.~P.,  2009, MNRAS, 398, 1698

\bibitem[\protect\citeauthoryear{{Schneider}, {Ehlers} \& {Falco}}{{Schneider}
  et~al.}{1995}]{Schneider1995book}
{Schneider} P.,  {Ehlers} J.,    {Falco} E.~E.,  1995, Gravitational Lenses.
Springer-Verlag

\bibitem[\protect\citeauthoryear{{Sharon}, {Gladders}, {Rigby}, {Wuyts},
  {Bayliss}, {Johnson}, {Florian} \& {Dahle}}{{Sharon}
  et~al.}{2014}]{Sharon2014}
{Sharon} K.,  {Gladders} M.~D.,  {Rigby} J.~R.,  {Wuyts} E.,  {Bayliss} M.~B.,
  {Johnson} T.~L.,  {Florian} M.~K.,    {Dahle} H.,  2014, ApJ, 795, 50

\bibitem[\protect\citeauthoryear{{Trotta}}{{Trotta}}{2008}]{BayesInTheSky2008Trotta}
{Trotta} R.,  2008, Contemporary Physics, 49, 71

\bibitem[\protect\citeauthoryear{{White}, {Canning}, {King}, {Lee}, {Russell},
  {Baum}, {Clowe}, {Coleman}, {Donahue}, {Edge}, {Fabian}, {Johnstone},
  {McNamara}, {O'Dea} \& {Sanders}}{{White} et~al.}{2015}]{white2146dynamics}
{White} J.~A.,  {Canning} R.~E.~A.,  {King} L.~J.,  {Lee} B.~E.,  {Russell}
  H.~R.,  {Baum} S.~A.,  {Clowe} D.~I.,  {Coleman} J.~E.,  {Donahue} M.,
  {Edge} A.~C.,  {Fabian} A.~C.,  {Johnstone} R.~M.,  {McNamara} B.~R.,
  {O'Dea} C.~P.,    {Sanders} J.~S.,  2015, MNRAS, 453, 2718

\bibitem[\protect\citeauthoryear{{Zitrin}, {Meneghetti}, {Umetsu},
  {Broadhurst}, {Bartelmann}, {Bouwens}, {Bradley}, {Carrasco}, {Coe}, {Ford},
  {Kelson}, {Koekemoer}, {Medezinski}, {Moustakas}, {Moustakas}, {Nonino},
  {Postman}, {Rosati}, {Seidel}, {Seitz}, {Sendra}, {Shu}, {Vega}, {Zheng}}{{Zitrin} et~al.}{2013}]{zitrin2013}
{Zitrin} A.,  {Meneghetti} M.,  {Umetsu} K.,  {Broadhurst} T.,  {Bartelmann}
  M.,  {Bouwens} R.,  {Bradley} L.,  {Carrasco} M.,  {Coe} D.,  {Ford} H.,
  {Kelson} D.,  {Koekemoer} A.~M.,  {Medezinski} E.,  {Moustakas} J.,
  {Moustakas} L.~A.,  {Nonino} M.,  {Postman} M.,  {Rosati} P.,  {Seidel} G.,
  {Seitz} S.,  {Sendra} I.,  {Shu} X.,  {Vega} J.,    {Zheng} W.,  2013, ApJ,
  762, L30  

\end{thebibliography}
\bibliographystyle{mn2e}

\appendix

\section{Multiple Image Listing}

\newpage

\begin{table}
\caption{A listing of multiple image systems near \bcga{}. There are four systems in this list.  All objects in a system are constrained to be at the same redshift.  For each element of a system the first letter of an image name corresponds to a multiple of a common source object. Images denoted with \dag{} are not included as constraints.  }
\label{tblMultipleImagesA}
\begin{tabular}{lccc}
	\hline
	System & Right Ascension & Declination & Notes \\
	\hline
	\multicolumn{4}{l}{System 1 `bra-ket'}\\
a1 & 239.04826 & 66.346907 & \\
a2 & 239.05165 & 66.349279 & \\
a3 & 239.06672 & 66.353407 & \\
b1 & 239.04864 & 66.347000 & \dag{} \\
b2 & 239.05165 & 66.349116 & \dag{} \\
b3 & 239.06710 & 66.353419 & \dag{} \\
c1 & 239.04886 & 66.347100 & \\
c2 & 239.05157 & 66.349027 & \\
c3 & 239.06730 & 66.353450 & \\
d1 & 239.04839 & 66.347198 & \\
d2 & 239.05131 & 66.349182 & \\
d3 & 239.06674 & 66.353526 & \\
	\multicolumn{4}{l}{System 2 }\\
e1 & 239.05054 & 66.347488 & \\
e2 & 239.05191 & 66.348565 & \\
e3 & 239.06695 & 66.353009 & \\
f1 & 239.05070 & 66.347593 & \\
f2 & 239.05175 & 66.348429 & \\
	\multicolumn{4}{l}{System 3 `mask'}\\
g1 & 239.06584 & 66.347623 & \\
g2 & 239.06673 & 66.348324 & \\
h1 & 239.06490 & 66.348031 & \\
h2 & 239.06563 & 66.348627 & \\
i1 & 239.06561 & 66.347423 & \dag{} \\
i2 & 239.06700 & 66.348519 & \dag{} \\
k1 & 239.06514 & 66.348038 & \dag{} \\
k2 & 239.06568 & 66.348510 & \dag{} \\
l1 & 239.06558 & 66.347902 & \dag{} \\
l2 & 239.06615 & 66.348362 & \dag{} \\
m1 & 239.06514 & 66.348038 & \\
m2 & 239.06568 & 66.348510 & \\
	\multicolumn{4}{l}{System 4}\\
j1 & 239.06186 & 66.347932 & \\
j2 & 239.06460 & 66.349884 & \\
j3 & 239.05053 & 66.344740 & \\
	\hline
\end{tabular}
\end{table}

\begin{table}
\caption{A listing of multiple image systems near \bcgb{}. All objects in a system are assumed to be at the same redshift.  For each element of a system the first letter of an image name corresponds to a multiple of a common source object. None of these systems are used as constraints in the final model. Objects denoted with $\diamond{}$ are listed here because colour, shape, location, and orientation make them interesting candidates for future models.    }
\label{tblMultipleImagesB}
\begin{tabular}{lccc}
	\hline
	System & Right Ascension & Declination & Notes \\
	\hline
	\multicolumn{4}{l}{System 5}\\
n1 & 238.99834 & 66.372898 & $\diamond{}$ \\
n2 & 238.99842 & 66.373224 & $\diamond{}$ \\
	\multicolumn{4}{l}{System 6}\\
p1 & 239.00083 & 66.370110 & $\diamond{}$ \\
p2 & 239.01642 & 66.372844 & $\diamond{}$ \\
q1 & 239.00073 & 66.370038 & $\diamond{}$ \\
q2 & 239.01659 & 66.372918 & $\diamond{}$ \\
	\multicolumn{4}{l}{System 7}\\
r1 & 239.01453 & 66.368228 & \\
r2 & 239.01499 & 66.368329 & \\
r3 & 239.01755 & 66.369324 & \\
s1 & 239.01438 & 66.368208 & \\
s2 & 239.01517 & 66.368373 & \\
s3 & 239.01739 & 66.369273 & \\
	\hline
\end{tabular}
\end{table}

\end{document}